\documentclass[nofootinbib,amsmath,amssymb,amsfonts,aps,prd,eqsecnum,superscriptaddress,reprint]{revtex4-1}

\pdfoutput=1
\usepackage{hyperref}
\usepackage{amsmath}
\usepackage{amssymb}
\usepackage{mathrsfs}
\usepackage{xcolor,graphicx}
\usepackage[caption=false]{subfig}
\usepackage[toc,page]{appendix}
\usepackage[section]{placeins}
\usepackage{breqn}
\usepackage[T1]{fontenc}
\usepackage[normalem]{ulem}
\usepackage{soul}

\makeatletter
\let\cat@comma@active\@empty
\makeatother

\begin{document}

\title{Consistent perturbative modeling of pseudo-Newtonian core-collapse supernova simulations}

\author{John Ryan Westernacher-Schneider}
\email{jwestern@email.arizona.edu}
\affiliation{Department of Astronomy/Steward Observatory, The University of Arizona, 933 N. Cherry Ave, Tucson, AZ 85721, USA}


\begin{abstract}
We write down and apply the linearized fluid and gravitational equations consistent with pseudo-Newtonian simulations, whereby Newtonian hydrodynamics is used with a pseudo-Newtonian monopole and standard Newtonian gravity for higher multipoles. We thereby eliminate the need to use mode function matching to identify the active non-radial modes in pseudo-Newtonian core-collapse supernova simulations, in favor of the less complex and less costly mode frequency matching method. In doing so, we are able to measure appropriate boundary conditions for a mode calculation.
\end{abstract}

\maketitle

\section{Introduction}
There is increasing attention to gravitational wave asteroseismology of core-collapse supernovae (CCSNe) from a theoretical perspective (eg.~\cite{murphy2009model, muller2013new, cerda2013gravitational, fuller2015supernova,torres2017towards,morozova2018gravitational, torres2018towards,westernacher2018turbulence,torres2019universal,vartanyan2019temporal, sotani2019dependence,westernacher2019multimessenger,warren2019constraining}). One challenge is identifying which hydrodynamical modes of the system are producing gravitational wave (GW) emission in simulations. This requires modeling in post-process. One strategy is to use simulation snapshots as background solutions for a perturbative mode calculation. Once the perturbative mode spectrum is obtained, a matching procedure is necessary to determine which modes are actually active in the simulation. A mode frequency matching procedure has been used frequently~\cite{torres2017towards,morozova2018gravitational,torres2018towards}, whereby the evolution of perturbative mode frequencies are overlaid on simulation gravitational wave spectrograms, and then matching is judged by frequency coincidence over time. 

However, some mode classes (particularly $p$-modes) tend to have frequencies which are roughly constant multiples of each other over time, with neighboring modes having frequencies being $\sim 5\,$-$10\%$ away. Frequency mismatches between simulations and perturbative calculations can arise due to the use of different equations of motion in the simulations versus those used in the perturbative calculation. For example, in~\cite{torres2017towards,torres2018towards} the general relativistic hydrodynamic equations were used in the perturbative calculation, with either no metric perturbations~\cite{torres2017towards} or a subset of possible metric perturbations~\cite{torres2018towards}. Their simulations correspondingly use general relativistic hydrodynamics and a spatially-conformally flat metric approximation for spacetime. As another example,~\cite{morozova2018gravitational} uses for their perturbative equations general relativistic hydrodynamics with either no metric perturbations or only lapse perturbations, supplemented with a Poisson equation to solve for the lapse perturbation. Their simulations on the other hand use Newtonian hydrodynamics and pseudo-Newtonian gravity. The ensuing frequency mismatches generated by the use of different equations may result in mode misidentification during a mode frequency matching procedure, particularly due the absence of the lapse function in the hydrodynamic fluxes in the simulations.

In~\cite{westernacher2018turbulence, westernacher2019multimessenger} a mode function matching procedure was followed instead. This entails comparing the mode functions computed perturbatively with the velocity data in the simulation. As in~\cite{morozova2018gravitational}, the simulations were pseudo-Newtonian, whereas the perturbative calculation used the general relativistic hydrodynamic equations in the Cowling approxmation (no metric perturbations), with the lapse function being the only non-zero metric component. The mode function matching procedure produced convincing mode identification despite the use of perturbative equations that are not consistent with the simulation, because neighboring mode functions have distinct enough morphology that the best-fitting mode function is clearly superior to the next-best-fitting one (provided the mode's excitation is large enough with respect to stochastic or nonlinear motions). A frequency mismatch between the best-fitting mode functions and the simulation frequencies of order $\sim 15\%$ was observed in~\cite{westernacher2018turbulence, westernacher2019multimessenger}, which is large enough to have caused a mode misidentification via mode frequency matching. During targeted modeling of the next galactic core-collapse supernova, this would have produced incorrect inferences about the source. Furthermore, mode misidentification in simulations can misinform analytic or semi-analytic modeling efforts of these systems.

However, mode function matching is considerably more complex and expensive than mode frequency matching. It is more complex because frequency masks have to be determined in order to apply appropriate spectral filtering on the velocity data from the simulation. It is more expensive because the entire fluid data in the system must be saved with sufficient temporal cadence such that the spectral resolution allows a clean Fourier extraction of individual mode activity. In~\cite{westernacher2018turbulence, westernacher2019multimessenger} axisymmetric simulations were performed, which alleviates the storage issue, but one wishes to identify modes in fully 3D simulations as well. Large searches of the CCSN progenitor parameter space would be hampered by the need to perform mode function matching. It would therefore be desirable to use the perturbative equations that are consistent with simulations, which, removing the need for the expensive mode function matching procedure.

In this work, we write down and apply the consistent linearized equations appropriate for pseudo-Newtonian codes such as~\texttt{PROMETHEUS/VERTEX}~\cite{rampp2002radiation, muller2010new, muller2012new, muller2013new, muller2014new},~\texttt{FLASH}~\cite{fryxell2000flash,dubey2009extensible},~\texttt{FORNAX}~\cite{skinner2019fornax},~\texttt{CHIMERA}~\cite{bruenn2018chimera}. As long as one does not solve for radial modes, these equations are simply the standard Newtonian ones. During testing we identify and correct a mistreatment of the boundary conditions~\cite{morozova2018gravitational, westernacher2018turbulence, westernacher2019multimessenger} for the gravitational potential perturbation. We are able to reproduce the quadrupolar mode frequencies of an equilibrium star evolved using \texttt{FLASH}. When applied to a CCSN simulation, we find the best-fitting mode functions have the correct frequency (i.e.~agreeing with the simulation) at the $2\%$ or sub-$1\%$ level, depending on the boundary conditions used. We also perform a residual test with the spherically-symmetric Euler equation, showing that the state of hydrostatic equilibrium (assumed in the perturbative calculation) is satisfied only at the $\sim 5\%$ level, whereas the terms coming from a time-dependent or non-steady ($v\neq 0$) background solution are negligible. This serves as a cautionary note for future applications of this perturbative modeling, but also suggests that including a time-dependent or non-steady background would not affect the calculation significantly. We find that the outer boundary condition on the fluid variables yielding the most precise matching with simulations (sub-$1\%$ level) is that of~\cite{torres2017towards}, where the radial displacement is taken to vanish at the shockwave location. The agreement is so striking that we are tempted to conclude that this is the physically correct boundary condition in the early post-bounce regime we are considering.

Note that the consistent perturbative modeling of pseudo-Newtonian simulations that we present here does \emph{not} answer the question of whether such simulations yield the correct mode excitation. Previously in~\cite{westernacher2018turbulence, westernacher2019multimessenger}, it was shown that, \emph{if} the perturbative modeling does not use the linearization of the equations being simulated, then mode function matching is necessary to correctly identify the active modes in a simulation. In this work, we simply use the consistent linearization to show that correct identification of active modes in a simulation is possible with mode frequency matching alone, and interesting physics can then be extracted (such as the physically correct boundary conditions for the perturbations). The question of whether the mode excitation itself is correct in pseudo-Newtonian simulations is left for future work. Previous studies indicate that mode frequencies are systematically shifted with respect to general relativity (see e.g.~\cite{mueller2008exploring}), and overestimated in particular~\cite{mueller2008exploring, westernacher2018turbulence, westernacher2019multimessenger}, but one cannot know for sure without directly identifying the excited modes in each case (e.g. by mode function matching). The pitfalls found in~\cite{westernacher2018turbulence, westernacher2019multimessenger} in using pseudo-Newtonian simulations to study mode frequencies were anticipated clearly in \cite{mueller2008exploring}.

We give a brief summary of the results of~\cite{westernacher2018turbulence, westernacher2019multimessenger} in Sec.~\ref{sec:summary}. We described our methods in Secs.~\ref{sec:methods} \& Appendix~\ref{sec:bcs}, and discuss our results in Sec.~\ref{sec:results}. Tests are presented in Appendix~\ref{sec:tests}. We use geometric units $G=c=1$ throughout, unless units appear explicitly.
%
%
\section{Simulations and background information} \label{sec:summary}
We analyze the non-rotating $20\,M_{\odot}$ zero-age main sequence mass CCSN progenitor presented previously in~\cite{westernacher2018turbulence, westernacher2019multimessenger}. It was simulated in axisymmetry using \texttt{FLASH}~\cite{fryxell2000flash,dubey2009extensible} until $\sim 100$ ms post-bounce. Mild excitation of hydrodynamic modes are excited at bounce, the amplitude of which is expected to be artificially enhanced due to asymmetries introduced during collapse by the cylindrical computational grid. However, the strength of excitation does not concern us here -- we simply seek to demonstrate mode identification. We defer to~\cite{westernacher2019multimessenger} for a more detailed description of the simulation details. We also defer details regarding the mode function matching method to~\cite{westernacher2018turbulence}, where they are described in the most depth. The method involves using spectrogram filter kernels to extract mode motions from the velocity data in the simulations, followed by vector spherical harmonic decompositions to extract the angular harmonic components. The resulting fields are then normalized before their overlaps with perturbative mode functions are computed.

Our main purpose here is to apply a consistent linear perturbative scheme to a snapshot from the simulation at $t\sim 40$ ms post-bounce, which was previously analyzed~\cite{westernacher2018turbulence, westernacher2019multimessenger}, to study multiple quadrupolar modes ($l=2,\, m=0$) of the system which are excited weakly at bounce. The first mode has a peak frequency of $515$ Hz\footnote{Note that the mode is described in~\cite{westernacher2019multimessenger} as having a frequency of $483$ Hz, which is the middle value of the spectrogram filter kernel used to extract it. However, $515$ Hz is the location of the peak Fourier amplitude in the GW signal.}. This mode was found in~\cite{westernacher2018turbulence, westernacher2019multimessenger} to have a radial order $n=4$, and we make the same conclusion here. The second quadrupolar mode we study has a less well-defined peak frequency (we estimate $1241$ Hz from the GW spectrum), and was not reported in~\cite{westernacher2018turbulence, westernacher2019multimessenger}. Note that due to an analysis error, perturbative mode frequencies in~\cite{westernacher2018turbulence} should be corrected by multiplying them by $\sim 1.5$.
%
%
\section{Perturbative scheme} \label{sec:methods}
We begin with the Newtonian perfect fluid and gravity equations,
\begin{eqnarray}
\partial_t \rho + \nabla_i \left( \rho v^i \right) &=& 0 \label{eq:restmass}\\
\partial_t \left( \rho v_i \right) + \nabla_j \left( \rho v^j v_i\right) + \partial_i P &=& - \rho \partial_i \Phi \label{eq:euler} \\
\nabla^2 \Phi &=& 4\pi \rho. \label{eq:poisson}
\end{eqnarray}
We linearize these equations with respect to a spherically symmetric equilibrium background solution, $\rho=\rho(r)$, $v^i(r) = (v^r(r),0,0)$, $P=P(r)$, $\Phi=\Phi(r)$, $\partial_r P/\rho = -\partial_r \Phi$. Denote Eulerian perturbations with $\delta$ and Lagrangian ones with $\Delta$, and substitute eg.~$\rho \rightarrow \rho + \delta \rho$ into Eqs.~\eqref{eq:restmass}-\eqref{eq:poisson}. Also use the condition of adiabatic perturbations coming from the energy equation,
\begin{eqnarray}
\frac{\Delta P}{\Delta \rho} = c_s^2 \label{eq:adiab}
\end{eqnarray}
where $c_s^2 = P\Gamma_1/\rho$ is the sound speed squared, $\Gamma_1$ is the adiabatic index for the perturbations, and eg.~$\Delta P = \delta P + \xi^i\nabla_i P$ where $\xi^i$ is the perturbative Eulerian fluid element displacement vector. The displacement vector $\xi^i$ is related to the velocity perturbation via $\delta v^i = \partial_t \xi^i + v^j\nabla_j \xi^i - \xi^j\nabla_j v^i$, which simplifies to $\delta v^i = \partial_t \xi^i$ when the background velocity is zero.

Linearization of Eqs.~\eqref{eq:restmass}-\eqref{eq:poisson} assuming axisymmetric perturbations $\xi^i = (\xi^r,\xi^{\theta},0)$ yields
\begin{eqnarray}
0 &=& \delta \rho + \rho \xi^i \partial_i \ln{\sqrt{\gamma}} + \rho \partial_i\xi^i + \xi^r\partial_r\rho \label{eq:restmasspert1}\\
0 &=& \partial_t^2 \xi^r + \frac{1}{\rho}\partial_r \delta P + \partial_r \delta \Phi - \frac{\delta\rho}{\rho^2}\partial_r P \label{eq:euler_rpert1}\\
0 &=& r^2 \partial_t^2 \xi^{\theta} + \frac{1}{\rho}\partial_{\theta} \delta P + \partial_{\theta} \delta\Phi \label{eq:euler_thpert1}\\
0 &=& \nabla^2 \delta\Phi - 4\pi \delta \rho \label{eq:poissonpert1}
\end{eqnarray}
where $\sqrt{\gamma} = r^2 \sin\theta$ is the square root of the flat 3-metric determinant in spherical coordinates. In deriving Eq.~\eqref{eq:restmasspert1} we integrated in time, setting the integration constant to zero~\cite{poisson2014gravity}. In Eq.~\eqref{eq:euler_thpert1} note the appearance of the factor $r^2$ in front of the time derivative, which comes from raising the index using the metric via $\partial_t^2 \xi_{\theta} = \gamma_{i\theta} \partial_t^2 \xi^i = \gamma_{\theta\theta} \partial_t^2 \xi^{\theta} = r^2 \partial_t^2 \xi^{\theta} $. Using the axisymmetric spherical harmonics $Y_l$ ($m=0$) and harmonic time dependence, we insert a separation of variables ansatz
\begin{eqnarray}
\delta\rho &=& \delta\hat{\rho}(r) Y_{l} e^{-i \sigma t} \nonumber\\
\delta P &=& \delta\hat{P}(r) Y_{l} e^{-i \sigma t} \nonumber\\
\delta\Phi &=& \delta\hat{\Phi}(r) Y_{l} e^{-i \sigma t} \nonumber\\
\xi^r &=& \eta_r(r) Y_{l} e^{-i\sigma t} \nonumber\\
\xi^{\theta} &=& \frac{\eta_{\perp}(r)}{r^2} \partial_{\theta} Y_{l} e^{-i \sigma t}.
\end{eqnarray}
We will assume $l\neq 0$. The angular frequency is $\sigma = 2 \pi f$. Note that we are using the coordinate basis $\{ (\partial_r)^i,(\partial_{\theta})^i,(\partial_{\phi})^i \}$ rather than the normalized coordinate basis $\{ \hat{\boldsymbol{r}}, \hat{\boldsymbol{\theta}}, \hat{\boldsymbol{\phi}} \}$, which explains the last ansatz having $\eta_{\perp}/r^2$ rather than $\eta_{\perp}/r$. Plugging these ansatz into Eq.~\eqref{eq:euler_thpert1} gives us a relation to eliminate $\delta\hat{P}$ via
\begin{eqnarray}
\delta\hat{P} = \rho \left(\sigma^2 \eta_{\perp} - \delta\hat{\Phi}\right).
\end{eqnarray}
The adiabatic condition then yields a relation which can be used to eliminate $\delta\hat{\rho}$ via
\begin{eqnarray}
\delta\hat{\rho} = \rho\left( \frac{\sigma^2}{c_s^2}\eta_{\perp} - \frac{\delta\hat{\Phi}}{c_s^2} - \mathcal{B}\eta_r \right),
\end{eqnarray}
where we have defined $\mathcal{B} \equiv \partial_r\ln\rho - (1/\Gamma_1)\partial_r \ln P$ as the Schwarzschild discriminant. In what follows, we also define $\tilde{G} \equiv \partial_r P/\rho = -\partial_r \Phi$, and the Brunt-V{\"a}is{\"a}l{\"a} frequency squared is $N^2 = \tilde{G}\mathcal{B}$. The linearization of the remaining Eqs.~\eqref{eq:restmasspert1} \&~\eqref{eq:euler_rpert1} \&~\eqref{eq:poissonpert1} yields
\begin{eqnarray}
0 &=& \partial_r \eta_r + \left[\frac{2}{r} + \frac{\partial_r P}{\Gamma_1 P}\right]\eta_r \nonumber\\
&\phantom{=}& \phantom{\partial_r \eta_r} + \left[\frac{\sigma^2}{c_s^2} - \frac{l(l+1)}{r^2}\right] \eta_{\perp} - \frac{1}{c_s^2}\delta\hat{\Phi} \label{eq:restmasspert2}\\
0 &=& \partial_r \eta_{\perp} - \left[1-\frac{N^2}{\sigma^2}\right] \eta_r + \mathcal{B}\eta_{\perp} - \frac{\mathcal{B}}{\sigma^2} \delta\hat{\Phi} \\
0 &=& \partial_r \delta\hat{\Phi} - F \\
0 &=& \partial_r F + \frac{2}{r}F + 4\pi\rho \mathcal{B} \eta_r - 4\pi\rho\frac{\sigma^2}{c_s^2} \eta_{\perp} \nonumber\\
&\phantom{=}& \phantom{\partial_r F + \frac{2}{r}F} + \left[ \frac{4\pi\rho}{c_s^2} - \frac{l(l+1)}{r^2} \right] \delta\hat{\Phi}, \label{eq:poissonpert2}
\end{eqnarray}
where we defined $F\equiv \partial_r \delta\hat{\Phi}$ to reduce the system to first order. In obtaining these equations we used the identity $\partial_{\theta}^2 Y_{l} + \cot\theta \partial_{\theta}Y_{l} = -l(l+1)Y_{l}$. Note these perturbative equations are the same equations as in~\cite{christensen1991solar} Eqs. (31-33), after changing the definitions $\delta\hat{\Phi} \leftrightarrow - \Phi^\prime$, $\xi_h \leftrightarrow \eta_{\perp}/r$. The latter identification comes both from different definitions of $\eta_{\perp}$ vs $\xi_h$ as well as the use of different basis vectors -- $\{ (\partial_r)^a, (\partial_\theta)^a, (\partial_{\phi})^a \}$ in our case vs $\{ \hat{\boldsymbol{r}}, \hat{\boldsymbol{\theta}}, \hat{\boldsymbol{\phi}} \}$ in~\cite{christensen1991solar}.

To solve these equations, we integrate from a small non-zero radius $r_0$ (typically $dr/5$ where $dr$ is the grid resolution), where we impose regularity conditions (see Appendix~\ref{sec:bcs}) in the form (assuming $l\neq 0$)
\begin{eqnarray}
\eta_r = A_0 r^{l-1}, &\;\;& \phantom{\partial_r} \eta_{\perp} = \frac{A_0}{l} r^{l\phantom{-1}} \nonumber\\
\delta\hat{\Phi} = C_0 r^{l\phantom{-1}}, &\;\;& \partial_r \delta\hat{\Phi} = l C_0 r^{l-1},
\end{eqnarray}
where $A_0$ is specified as a small number ($10^{-5}$ in our case) which encodes the overall amplitude of the perturbation, and $C_0$ is searched for via a root-finding algorithm such that an outer boundary condition on $\delta\hat{\Phi}$ is satisfied -- see Appendix~\ref{sec:bcs} for a detailed description. This outer boundary condition on $\delta\hat{\Phi}$ was not imposed in~\cite{morozova2018gravitational}, where instead $\delta\hat{\Phi}\vert_{r_0} = 0 = \partial_r \delta \hat{\Phi}\vert_{r_0}$ was used. This error was repeated in subsequent work, including~\cite{westernacher2018turbulence, radice2019characterizing, westernacher2019multimessenger}, but does not affect any of the results obtained in the Cowling approximation.

We validate our current Newtonian perturbative scheme on a Newtonian polytropic star in Appendix~\ref{sec:tests}, and demonstrate that the effect of ignoring the outer boundary condition on $\delta\hat{\Phi}$ is large mode frequency errors for modes of low radial order. 

We also demonstrate in Appendix~\ref{sec:tests} that our current Newtonian perturbative scheme recovers the non-radial modes of equilibrium stars evolved in a pseudo-Newtonian system using \texttt{FLASH}. This system has a phenomenologically modified monopole gravitational potential designed to mimic relativistic stars (\cite{marek2006exploring} Case A). This demonstrates that we can solve for non-radial modes even though we do not have an equation of motion for the monopole potential. Such an equation never appears in our derivation above, because we assumed $l\neq 0$.

Having the consistent perturbative scheme for such pseudo-Newtonian simulations allows us to investigate how well other aspects of the approximation (the assumption of equilibrium background, zero background velocity, and spherical averaging) actually affect the mode identification.

The other outer boundary condition concerning the fluid variables is considerably more uncertain. In~\cite{morozova2018gravitational} it was taken to be $\Delta P \vert_{r=R} = 0$ for some outer boundary $R$ representing the proto-neutron star (PNS) surface, and in~\cite{torres2017towards} was taken to be $\eta_r\vert_{\mathrm{shockwave}}=0$. With the consistent perturbative equations, we can instead simply plug in the frequency observed in the simulation and see whether the resulting mode function matches the simulated velocity data well. We can also try to infer an appropriate outer boundary condition on the fluid variables in this way. Thus, we can turn the problem around and attempt to \emph{measure} the appropriate boundary condition. Theoretically, the boundary condition must account for the Rankine-Hugoniot jump conditions across the accretion shock, which in turn depend upon the state of the supersonically accreting material upstream from the shockwave (see e.g.~\cite{foglizzo2007instability, laming2007analytic}).

\begin{figure}
\centering
\includegraphics[width=0.48\textwidth]{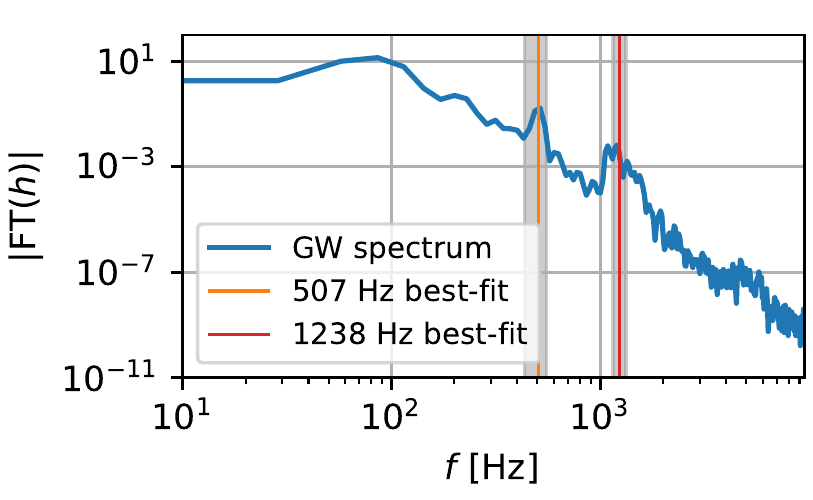}

\caption{Normalized GW spectrum averaged over $t\in [30,50]$ ms post-bounce, computed using a Bohman window with 35 ms width. Two frequencies of the best-fit mode functions are indicated at 507 Hz and 1238 Hz, corresponding to weakly excited quadrupolar modes. These compare well with the peaks in the GW spectrum at $515$ and $1241$ Hz. The shaded areas indicate the frequency extent of the spectral filter used in~\cite{westernacher2018turbulence, westernacher2019multimessenger} to extract the velocity data, against which perturbative mode functions are matched.} \label{fig:gwspec}
\end{figure}

%
%

%
%
\section{Results} \label{sec:results}

We show the GW spectrum in Fig.~\ref{fig:gwspec}, which is computed using a Bohman window with 35 ms width, and averaged over times $t\in [30,50]$ ms. The grey shaded intervals indicate the frequency extent of the spectral filters used to extract the velocity data from the simulation. A snapshot of that data near $t=40$ ms is then matched with perturbative solutions, with the frequency as the free parameter in the perturbative solutions. The perturbative solutions whose modefunction matches the velocity data best have frequencies of $507$ and $1238$ Hz, which compares well with the peaks in Fig.~\ref{fig:gwspec}.

\begin{figure}
\centering
\includegraphics[width=0.48\textwidth]{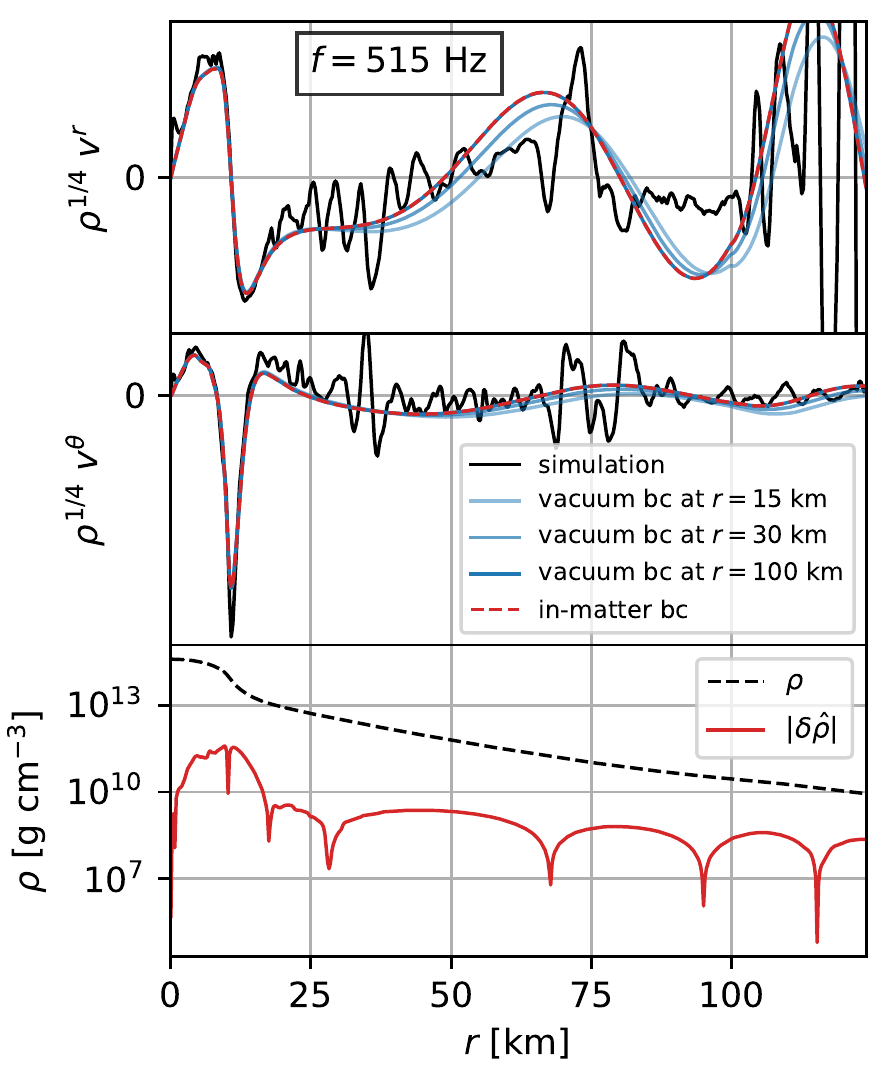}

\caption{Upper two panels: Normalized perturbative solutions plotted on a linear vertical scale, with frequency corresponding to the simulation, $f=515$ Hz, for varying outer boundary condition on the Newtonian potential perturbation. The vacuum boundary condition Eq.~\eqref{eq:phiobdy} is imposed at $r_{\mathrm{bc}}=\{15,30,100\}\,$km, and is seen to approach the in-matter boundary condition case~\eqref{eq:phiobdy_full} as the boundary is placed farther out. The perturbative solutions are a poor representation of the simulation beyond $\sim 10$ km. Bottom panel: The density $\rho$ and density perturbation $\vert \delta \hat{\rho} \vert$ are displayed for reference. The density perturbation becomes negligible beyond $\sim 60$ km. The shockwave is located at $r\sim 125$ km at this time $40$ ms post-bounce.} \label{fig:515Hz}
\end{figure}

Our first finding is that plugging in the simulation frequency $f\sim 515$ Hz (disregarding any outer boundary condition on the fluid variables) yields a perturbative solution that fits the simulation data well -- see Fig.~\ref{fig:515Hz}. In the top two panels we show the $515$ Hz perturbative solution (weighted by $\rho^{1/4}$) for various boundary conditions on $\delta\hat{\Phi}$, namely the vacuum one (Eq.~\eqref{eq:phiobdy}) imposed at various radii, as well as the in-matter one (Eq.~\eqref{eq:phiobdy2}) which does not depend on the outer boundary location. Note we plot on an arbitrary linear vertical scale. The result obtained using the vacuum boundary condition approaches the in-matter one rapidly as the outer boundary moves out, because the density perturbation $\delta\hat{\rho}$ becomes negligible for $r\gtrsim 60$ km (see bottom panel). For the rest of our results we use the in-matter boundary condition Eq.~\eqref{eq:phiobdy2}.

\begin{figure}
\centering
\includegraphics[width=0.48\textwidth]{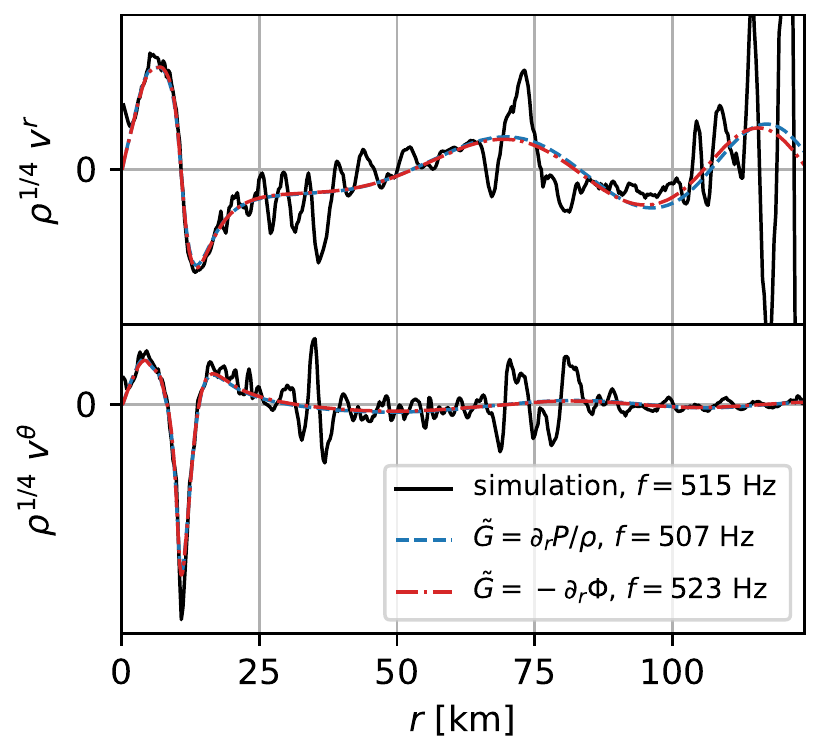}

\caption{The best-fit perturbative solutions for two different choices $\tilde{G}=\partial_r P/\rho$ and $\tilde{G}=-\partial_r \Phi$, which in a true spherically symmetric equilibrium would yield the same result. These choices yield frequencies of $507$ Hz and $523$ Hz, respectively. This is a mistmatch with the simulation frequency $515$ Hz by $\pm$1.6\%. These perturbative solutions have radial order $n=4$ if counted up to the shockwave location $r=125$ km.} \label{fig:515Hz_bfm}
\end{figure}

Next we do a search over frequency (again disregarding outer boundary conditions for the fluid variables) to find the best-fitting perturbative solution to the simulation data. The fit quality is computed by normalizing the $\sqrt{\rho}$-weighted velocities and computing a Frobenius norm of their difference (see~\cite{westernacher2019multimessenger}). The result is shown in Fig.~\ref{fig:515Hz_bfm}. Despite not smoothing the simulated data, the agreement is nonetheless striking. We again weight the velocity by $\rho^{1/4}$ to allow easier visual inspection (compared to a $\sqrt{\rho}$-weighting). We stress that this is an unforgiving way of displaying the agreement. The radial nodes of the best-fit perturbative solution are consistent with those found in~\cite{westernacher2018turbulence, westernacher2019multimessenger}, i.e. $n=4$ when counted within the shockwave (which is located at $r\sim 125$ km at this snapshot). Note that since our background is not actually in equilibrium, we have an ambiguity in how we apply the perturbative scheme. Namely, we can set $\tilde{G} = \partial_r P/\rho$ or $\tilde{G} = -\partial_r \Phi$\footnote{This is not the only ambiguity. Wherever a pressure gradient or gravitational potential gradient appears, one could switch it out with the other using $\partial_r P = -\rho\partial_r \Phi$.}. We show both cases in Fig.~\ref{fig:515Hz_bfm}, which yield best-fit solutions with frequencies of $507$ Hz and $523$ Hz, respectively. Both choices are equally accurate for this mode, but unless otherwise specified we will use $\tilde{G} = \partial_r P/\rho$.

\begin{figure}
\centering
\includegraphics[width=0.48\textwidth]{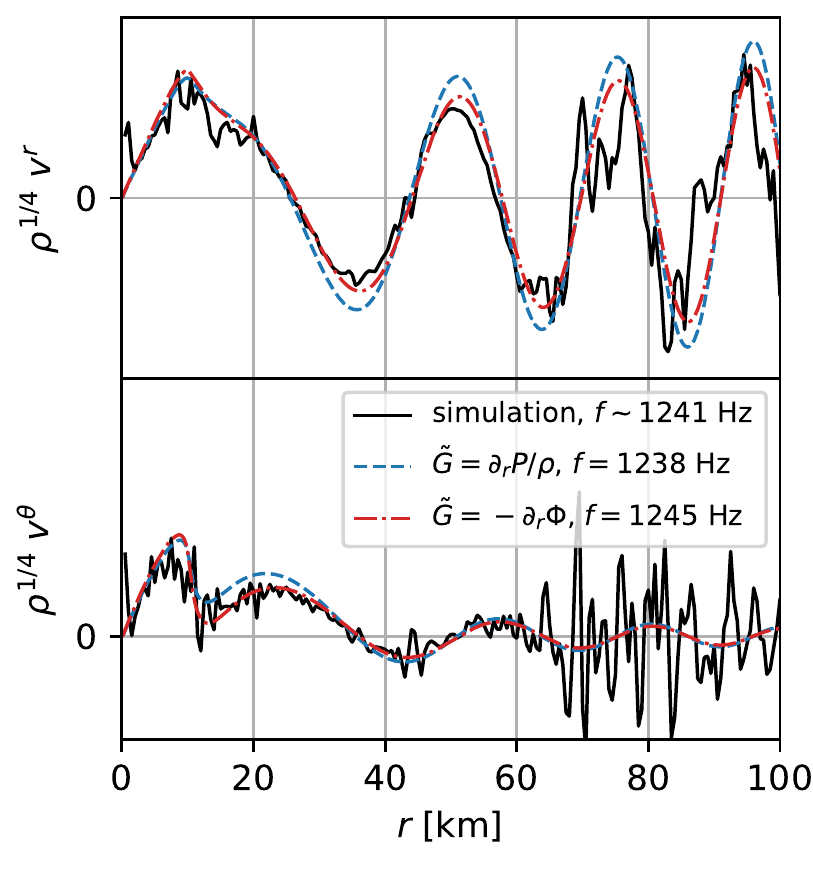}

\caption{Same as Fig.~\ref{fig:515Hz_bfm} but for the $\sim 1240$ Hz peak.} \label{fig:1240Hz_bfm}
\end{figure}

In Fig.~\ref{fig:1240Hz_bfm} we show the analogous plot for the $1241$ Hz frequency mode, showing a similar level of agreement. The best-fitting perturbative solutions have frequencies of $1238$ and $1245$ Hz for the cases $\tilde{G}=\partial_r P/\rho$ and $\tilde{G}=-\partial_r \Phi$, respectively. This is $0.24\%$ and $0.32\%$ disagreement, respectively.

We now reinstate outer boundary conditions for the fluid variables. Our purpose is to ``measure" the boundary conditions which will yield a mode function spectrum such that the best-fit mode function has a frequency which is (at least similar to) the simulation. If such a boundary condition existed, then one could safely identify modes in pseudo-Newtonian simulations by doing frequency matching alone, removing the need for the complicated and expensive mode function matching procedure described in~\cite{westernacher2018turbulence, westernacher2019multimessenger}. 

\begin{figure}
\centering
\includegraphics[width=0.48\textwidth]{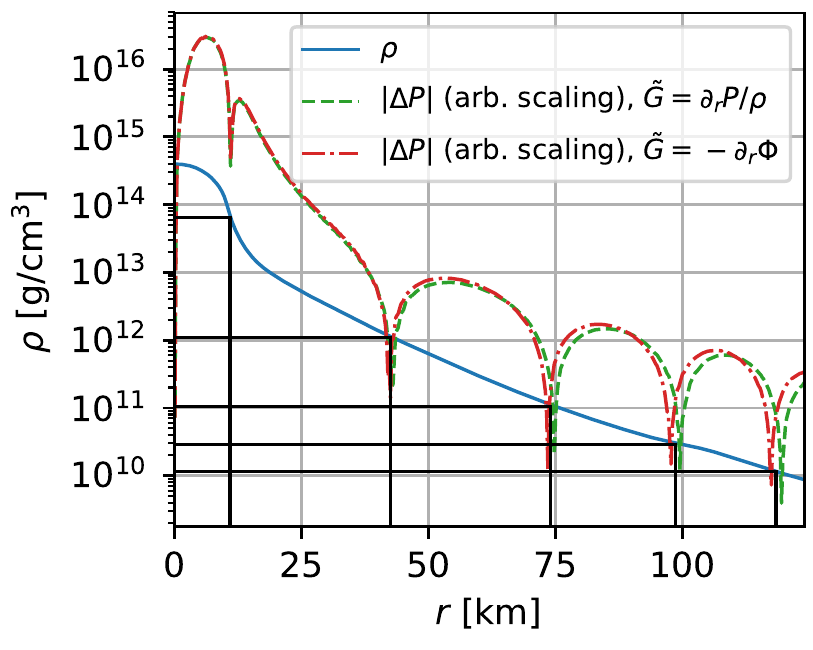}

\caption{Lagrangian pressure perturbation $\vert \Delta P\vert$ corresponding to the best-fit perturbative solutions in Fig.~\ref{fig:515Hz_bfm} displayed on an arbitrary logarithmic scale. The rest mass density of the spherically-averaged background is also displayed on an accurate logarithmic scale. Zeros of the Lagrangian pressure perturbation are indicated, which suggest appropriate values of $\rho$ at which $\Delta P = 0$ should be enforced during a mode search.} \label{fig:515Hz_DeltaP}
\end{figure}

\begin{figure}
\centering
\includegraphics[width=0.48\textwidth]{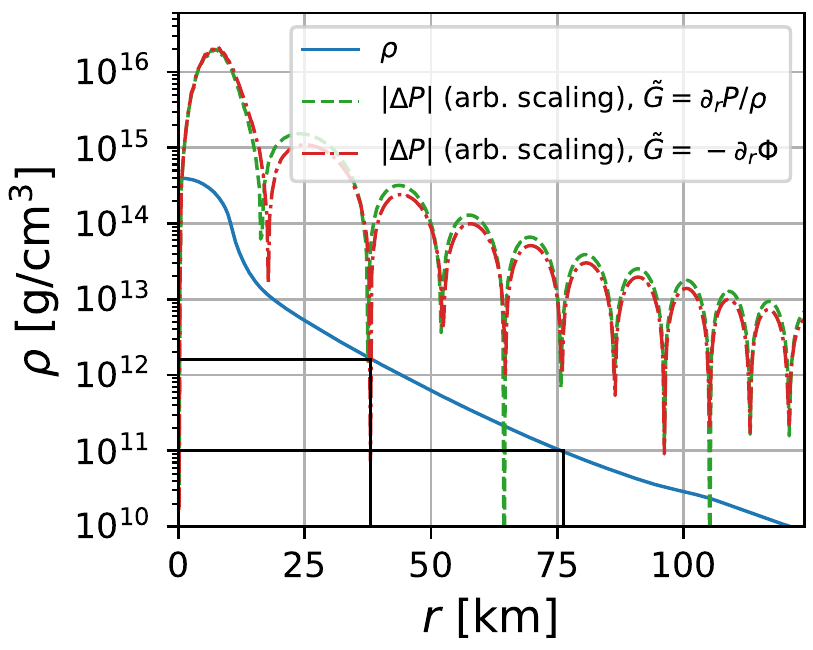}

\caption{Same as Fig.~\ref{fig:515Hz_DeltaP} but for the $\sim 1240$ Hz mode. The zeros of $\Delta P$ occuring near $\rho = 10^{12},10^{11}$ g$\,$cm$^{-3}$ are shown.} \label{fig:1240Hz_DeltaP}
\end{figure}

In Fig.~\ref{fig:515Hz_DeltaP} we plot the absolute value of the Lagrangian pressure perturbation corresponding to the best-fitting perturbative solutions for the $515$ Hz mode in Fig.~\ref{fig:515Hz_bfm} on an arbitrary logarithmic scale. The analogous plot for the $1241$ Hz mode is displayed in Fig.~\ref{fig:1240Hz_DeltaP}. The Lagrangian pressure perturbation is overlaid on the background density profile, which is plotted on a faithful logarithmic scale. We indicate the location of the zero-crossings of $\Delta P$ with dotted lines, and also indicate the corresponding density value there. Zero-crossings for the $515$ Hz case occur near $\{6\times 10^{13},10^{12},10^{11},10^{10}\}$ g$\,$cm$^{-3}$. A common definition for the PNS surface is e.g.~$\rho=10^{11}\,$g$\,$cm$^{-3}$, and a zero-crossing at that location also occurs for the $1241$ Hz mode in Fig.~\ref{fig:1240Hz_DeltaP}. These zero-crossings are not enforced, and if they are not mere coincidences then they could be physically meaningful if they work for different modes.

In Tables~\ref{table515} \&~\ref{table1240}, for various outer boundary conditions on the fluid variables we show the mode properties with nearest and next-nearest frequencies to the simulation (subscripts $_{\mathrm{best}}$ and $_{\mathrm{next}}$, respectively). All choices listed, aside from $\Delta P \vert_{\rho=10^{12}}=0$ which fails to reproduce the $1241$ Hz mode, yield a clear relative distinction between the best-fit and the next-best one, and could therefore be regarded as safe to use during a mode frequency matching procedure. However, the boundary condition of~\cite{torres2017towards}, $\eta_r \vert_{\mathrm{shockwave}}=0$, yields remarkable sub-$1\%$ agreement for both modes, suggesting it is the physically correct one in this regime.

\begin{table}[]
\begin{tabular}{c|c|c|c|c|}
Bdy. condition    & $f_{\mathrm{best}}$ [Hz], diff & $n_{\mathrm{best}}$ & $f_{\mathrm{next}}$ [Hz], diff & $n_{\mathrm{next}}$ \\\hline
$\Delta P\vert_{\rho\sim 10^{12}}=0$   & 504, -2.1\% & 4 & 381, -26\% & 4 \\
$\Delta P\vert_{\rho\sim 10^{11}}=0$   & 504, -2.1\% & 4 & 436, -15\% & 5 \\
$\Delta P\vert_{\rho\sim 10^{10}}=0$   & 503, -2.3\% & 4 & 463, -10\% & 5 \\
${\eta_r}\vert_{\mathrm{shockwave}}=0$ & 513, -0.4\% & 4 & 491, -4.7\% & 5
\end{tabular}
\caption{Modes with nearest ($f_{\mathrm{best}}$) and next-nearest ($f_{\mathrm{next}}$) frequencies to the simulation value of $515$ Hz, for varying boundary conditions. We use $\tilde{G}=\partial_r P/\rho$. The subscripts on $\Delta P$ (eg.~$\Delta P\vert_{\rho\sim 10^{12}}$) indicate density in units of g$\,$cm$^{-3}$. The nearest modes are highlighted in bold.} \label{table515}
\end{table}

\begin{table}[]
\begin{tabular}{c|c|c|c|c|}
Bdy. condition    & $f_{\mathrm{best}}$ [Hz], diff & $n_{\mathrm{best}}$ & $f_{\mathrm{next}}$ [Hz], diff & $n_{\mathrm{next}}$ \\\hline
$\Delta P\vert_{\rho\sim 10^{12}}=0$   & 1101, -11\% & 8 & 1532, +23\% & 12 \\
$\Delta P\vert_{\rho\sim 10^{11}}=0$   & 1239, -0.2\% & 9 & 1073, -14\% & 8 \\
$\Delta P\vert_{\rho\sim 10^{10}}=0$   & 1235, -0.5\% & 9 & 1357, +9.3\% & 10 \\
${\eta_r}\vert_{\mathrm{shockwave}}=0$ & 1248, -0.6\% & 9 & 1137, -8.4\% & 8
\end{tabular}
\caption{Same as Table~\ref{table515} but for the $\sim 1240$ Hz mode. Nodes are counted up to the shockwave at $r=125$ km.} \label{table1240}
\end{table}

\section{Outlook and conclusions} \label{ch:CCSNconc}

In this work, we presented and tested perturbative equations which are the consistent linear approximation of pseudo-Newtonian systems whereby one uses Newtonian hydrodynamics, standard Newtonian gravity for non-radial components of the potential, and some non-standard monopole potential such as that of~\cite{marek2006exploring} Case A. This system of equations allows one to solve for non-radial modes, thereby allowing identification of active modes in pseudo-Newtonian simulations (eg.~\texttt{PROMETHEUS/VERTEX}~\cite{rampp2002radiation, muller2010new, muller2012new, muller2013new, muller2014new},~\texttt{FLASH}~\cite{fryxell2000flash,dubey2009extensible},~\texttt{FORNAX}~\cite{skinner2019fornax},~\texttt{CHIMERA}~\cite{bruenn2018chimera}) using mode frequency matching. This alleviates the need to perform the complex and expensive mode function matching procedure of~\cite{westernacher2018turbulence, westernacher2019multimessenger}. 

We found that the imposing vanishing radial displacement as an outer boundary condition (as in~\cite{torres2017towards}) yields remarkable sub-$1\%$ agreement between perturbative mode frequencies and the simulation, suggesting that this is the physically correct choice. However, imposing a vanishing Lagrangian pressure perturbation at the radii where $\rho=\{10^{11},10^{10}\}$ g$\,$cm$^{-3}$ (the last value being used in~\cite{morozova2018gravitational}) should also prevent mode misidentification. These conclusions ought to be tested in other regimes, eg.~later times $t>100$ ms and different progenitor stars.

\acknowledgements
We thank Evan O'Connor for comments and insight regarding neutrino pressure gradients, and both Evan O'Connor and Sean M. Couch for \texttt{FLASH} code development and running the simulations analyzed in this work. We also thank an anonymous referee for providing a deeper context of this work within the existing literature. This research was supported by National Science Foundation Grant No. PHY-1912619 at the University of Arizona.

Software: Matplotlib \cite{Hunter:2007}, FLASH \cite{fryxell2000flash,dubey2009extensible,couch13, o2018two}, SciPy \cite{scipy}.

%
%
\appendix
\section{Boundary conditions} \label{sec:bcs}
In this section we give details of how boundary conditions are derived, for the purpose of being pedagogical. We use the strategy of~\cite{hurley1966oscillations}, except applied directly to our equations~\eqref{eq:restmasspert2}-\eqref{eq:poissonpert2}.

We wish to determine the behavior of $\{\eta_r,\eta_{\perp},\delta \hat{\Phi}\}$ in a neighborhood of the origin $r=0$. For this purpose, we make the ansatz 
\begin{eqnarray}
\eta_r &=& r^a \sum_{n=0}^{\infty}{A_n r^n} \nonumber\\
\eta_{\perp} &=& r^b \sum_{n=0}^{\infty}{ B_n r^n} \nonumber\\ 
\delta \hat{\Phi} &=& r^c \sum_{n=0}^{\infty}{ C_n r^n}, \nonumber
\end{eqnarray}
where $A_n, B_n, C_n$ are constant coefficients nonzero when $n=0$ (do not confuse $n$ in this context with the radial order of modes), and $a,b,c$ are constant exponents to be determined. We require $
a,b,c\geq 0$ by regularity at the origin. This ansatz is a generalization of the Frobenius method to a system of equations. The derivatives we need are
\begin{eqnarray}
\partial_r \delta \hat{\Phi} &=& r^c \sum_{n=0}^{\infty}{(n+c)C_n r^{n-1}} \\
\partial_r^2 \delta \hat{\Phi} &=& r^c \sum_{n=0}^{\infty}{(n+c)(n+c-1)C_n r^{n-2}},
\end{eqnarray}
and similar expressions for $\partial_r \eta_r,\partial_r \eta_{\perp}$.

Plugging these ansatz into our equations~\eqref{eq:restmasspert2}-\eqref{eq:poissonpert2} and collecting terms proportional to $r^a,r^b,r^c$, we schematically obtain
\begin{eqnarray}
0 &=& Q_a r^a + Q_b r^b + Q_c r^c \nonumber\\
0 &=& R_a r^a + R_b r^b + R_c r^c \\ \label{eq:bdyschem}
0 &=& S_a r^a + S_b r^b + S_c r^c,\nonumber
\end{eqnarray}
where the coefficients are
\begin{eqnarray}
Q_a &=& \sum_{n=0}^{\infty}{nA_nr^{n-1}} + \left[\frac{2+a}{r} + \frac{\partial_r P}{\Gamma_1 P} \right] \sum_{n=0}^{\infty}{A_n r^n} \nonumber\\
Q_b &=& \left[\frac{\sigma^2}{c_s^2} - \frac{l(l+1)}{r^2}\right] \sum_{n=0}^{\infty} B_n r^n \nonumber\\
Q_c &=& -\frac{1}{c_s^2} \sum_{n=0}^{\infty}{C_n r^n} \nonumber\\
R_a &=& -\left[1-\frac{N^2}{\sigma^2}\right] \sum_{n=0}^{\infty}{A_n r^n} \nonumber\\
R_b &=& \sum_{n=0}^{\infty}{nB_n r^{n-1}} + \left[ \frac{b}{r} + \mathcal{B} \right] \sum_{n=0}^{\infty}{B_n r^n} \nonumber\\
R_c &=& -\frac{\mathcal{B}}{\sigma^2} \sum_{n=0}^{\infty}{C_n r^n} \nonumber\\
S_a &=& 4\pi\rho \mathcal{B} \sum_{n=0}^{\infty}{A_n r^n} \nonumber\\
S_b &=& -4\pi\rho \frac{\sigma^2}{c_s^2} \sum_{n=0}^{\infty}{B_n r^n} \nonumber\\
S_c &=& \sum_{n=0}^{\infty}{n^2 C_n r^{n-2}} + \left[ \frac{2c+1}{r} \right] \sum_{n=0}^{\infty}{nC_n r^{n-1}} \nonumber\\
&+& \left[\frac{c(c+1)-l(l+1)}{r^2} + \frac{4\pi\rho}{c_s^2}\right] \sum_{n=0}^{\infty}{C_n r^n}.
\end{eqnarray}
Since Eqs.~\eqref{eq:bdyschem} hold in a neighborhood of the origin, the full coefficients in front of each power of $r$ (once collected) must vanish independently. We are interested in the vanishing of the lowest order terms.

In the Frobenius method, only one equation is being solved. This means only one unknown exponent (eg. $a$ above) appears in the equation once the ansatz is plugged in. This makes identifying orders in $r$ straightforward. In our case, we have a system of equations and multiple unknown exponents $a,b,c$ appear in each equation. This makes identifying orders in $r$ more complicated, but we can proceed by considering all possible cases and systematically eliminating them. This is what we do next.

Since we are interested in the lowest nontrivial order, it suffices to truncate every sum after the first nonzero term. We also need to consider the order carried by the background quantities. In particular, since the pressure and density are spherically-symmetric quantities with even parity, we have $P \simeq P\vert_{0} + P^{\prime\prime} r^2/2$ and $\rho \simeq \rho\vert_{0} + \rho^{\prime\prime} r^2/2$, where we use a double prime superscript to denote a second radial derivative evaluated at the origin, to avoid cumbersome notation. This means $\partial_r P = P^{\prime\prime} r = \mathcal{O}(r)$ and $\partial_r \rho = \rho^{\prime\prime} r = \mathcal{O}(r)$. Thus $\mathcal{B} = \partial_r\rho/\rho - \partial_r P/(\Gamma_1 P) \simeq [\rho^{\prime\prime}/\rho - P^{\prime\prime}/(\Gamma_1 P)] r = \mathcal{O}(r)$. Similarly, $\tilde{G} = \partial_r P/\rho \simeq P^{\prime\prime} r/\rho = \mathcal{O}(r)$, and so by extension $N^2 = \tilde{G} \mathcal{B} = \mathcal{O}(r^2)$.  Inserting these expansions into Eqs.~\eqref{eq:bdyschem} and keeping lowest-order terms for each of the $r^a,r^b,r^c$ terms separately, we obtain
\begin{eqnarray}
0 &=& (2+a)A_0 r^{a-1} - B_0 l(l+1) r^{b-2} - \frac{C_0}{c_s^2} r^c \label{eq:low1} \\
0 &=& -A_0 r^a + B_0 b r^{b-1} - \frac{C_0}{c_s^2} \left[ \frac{\rho^{\prime\prime}}{\rho} - \frac{P^{\prime\prime}}{\Gamma_r P} \right] r^{c+1} \label{eq:low2} \\
0 &=& 4\pi\rho\left[\frac{\rho{\prime\prime}}{\rho} - \frac{P{\prime\prime}}{\Gamma_1 P}\right] A_0 r^{a+1} - 4\pi\rho \frac{\sigma^2}{c_s^2} B_0 r^b \nonumber\\
&+& \left[ c(c+1) - l(l+1) \right] C_0 r^{c-2}. \label{eq:low3}
\end{eqnarray}

At this stage we do not know whether we have kept consistent orders in $r$, since we do not know the relationship between the exponents $a,b,c$. However, when considering Eq.~\eqref{eq:low3}, notice that the exponents will not depend upon the background solution if and only if the $r^{c-2}$ term is the lowest order one. Independence from the background solution is a property we desire\footnote{Although it would be interesting to know whether ``special" perturbations of stars with exponents depending upon the background solution are ever relevant in practice.}, thus we demand that the $r^{c-2}$ term must vanish, i.e.~$c=l$. This also implies $c-2 < a+1$ and $c-2 < b$.

The same consideration applied to Eq.~\eqref{eq:low1} means that one or both of the $r^{a-1}$ and $r^{b-2}$ terms must be lowest order. If the $r^{b-2}$ term is lowest order by itself, that implies $l=0$. If we are not interested in radial modes (in this work, we are not), then we can discard this possibility. On the other hand, if the $r^{a-1}$ term is lowest order by itself, that implies $a=-2$ which would violate regularity at the origin. Thus we must conclude that both terms are lowest order, i.e.~$a=b-1$ and $(2+a)A_0 = B_0 l(l+1)$. 

Lastly, consider Eq.~\eqref{eq:low2}. If the exponents are to be independent of the background quantities, then one or both of the $r^a$ and $r^{b-1}$ terms must be lowest order. But we already established that $a=b-1$, thus they are both lowest order. This yields $A_0 = b B_0$. Combining this relation with the one obtained previously from Eq.~\eqref{eq:low1} and using $a=b-1$, we finally find
\begin{eqnarray}
a=l-1, \;\; b=l, \;\; c=l.
\end{eqnarray}
Therefore, in a neighborhood of the origin,
\begin{eqnarray}
\eta_r = A_0 r^{l-1}, &\;\;& \phantom{\partial_r} \eta_{\perp} = \frac{A_0}{l} r^{l\phantom{-1}} \nonumber\\
\delta\hat{\Phi} = C_0 r^{l\phantom{-1}}, &\;\;& \partial_r \delta\hat{\Phi} = l C_0 r^{l-1}. \label{eq:origin}
\end{eqnarray}
Beware that we are not using the normalized coordinate basis. In the normalized basis, one instead has $\eta_{\perp} = (A_0/l) r^{l-1}$.

In the numerical integration, we begin a small distance away from the origin (eg.~$dr/5$, where $dr$ is the grid resolution) and use Eqs.~\eqref{eq:origin} as initial conditions. This requires specification of $A_0, C_0$ and the angular frequency $\sigma$. The choice of $A_0$ amounts to an arbitrary amplitude, which we choose to be $A_0 = 10^{-5}$. 

For each value of angular frequency $\sigma$, we perform a root-finding procedure to converge upon the value of $C_0$ such that at the outer boundary $r=R$ we have~\cite{christensen1991solar}
\begin{eqnarray}
\left[\partial_r \delta\hat{\Phi} + \frac{l+1}{r} \delta \hat{\Phi}\right]\vert_{r=R} = 0. \label{eq:phiobdy}
\end{eqnarray}
This relation can be derived from the solution for the $l$th spherical harmonic moment of the Poisson equation~\cite{poisson2014gravity}
\begin{eqnarray}
\delta\hat{\Phi} = -\frac{4\pi}{2l+1}\frac{1}{R^{l+1}} \int_0^r{\delta \hat{\rho}(\tilde{r}) \tilde{r}^{l+2}d\tilde{r}}, \label{eq:phiobdy_full}
\end{eqnarray}
valid when $\delta \hat{\rho}(r) = 0$ for $r>R$. In the case of our CCSN system, $l$th moment rest mass perturbations $\delta \hat{\rho}$ likely escape out through $r=R$, but to the extent that it is of small amplitude and leaks into different harmonics $l^\prime\neq l$, it can be ignored. If it cannot be ignored, then one should instead integrate the perturbative system beyond $r=R$ and then impose
\begin{eqnarray}
\!\!\!\!\!\!\!\!\left[\partial_r \delta\hat{\Phi} + \frac{l+1}{r} \delta \hat{\Phi}\right]\vert_{r=R} = -4\pi R^{l-1}\!\!\! \int_R^{\infty}{\frac{\delta\hat{\rho}}{r^{l-1}}dr}, \label{eq:phiobdy2}
\end{eqnarray}
where the infinite upper limit of integration is understood to be replaced by an appropriate outermost radius, eg. the grid boundary or the CCSN shockwave. When using Eq.~\eqref{eq:phiobdy2}, one must integrate past $R$ in order to obtain $\delta\hat{\rho}$ over the domain of interest. The choice of $R$ is irrelevant. Note that
\begin{eqnarray}
\delta \hat{\rho} = \rho \left( \frac{\sigma^2}{c_s^2} \eta_{\perp} - \frac{\delta\hat{\Phi}}{c_s^2} - \mathcal{B}\eta_r \right). \label{eq:deltarho}
\end{eqnarray}

Also, it is advisable to enforce Eq.~\eqref{eq:phiobdy} at the outer boundary rather than Eq.~\eqref{eq:phiobdy_full}, in order to get control of the first derivative $\partial_r \delta\hat{\Phi}$.

The root-finding loop for $C_0$ is nested inside a root-finder for the angular frequency $\sigma$, which yields either vanishing Lagrangian pressure perturbation at the outer boundary 
\begin{eqnarray}
\Delta P\vert_{R} = [\rho \sigma^2 \eta_{\perp} - \rho \delta\hat{\Phi} + \eta_r \partial_r P]\vert_{R} = 0, \label{eq:DeltaP0}
\end{eqnarray}
corresponding to a free surface, or vanishing radial displacement 
\begin{eqnarray}
\eta_r\vert_{R} = 0, \label{eq:etar0}
\end{eqnarray}
depending on one's choice.
%
%

\section{Tests of perturbative scheme} \label{sec:tests}

In this section we demonstrate the accuracy of our mode solver on both a Newtonian polytropic star and a pseudo-Newtonian ``TOV'' star.

\begin{figure}
\centering
\includegraphics[width=0.48\textwidth]{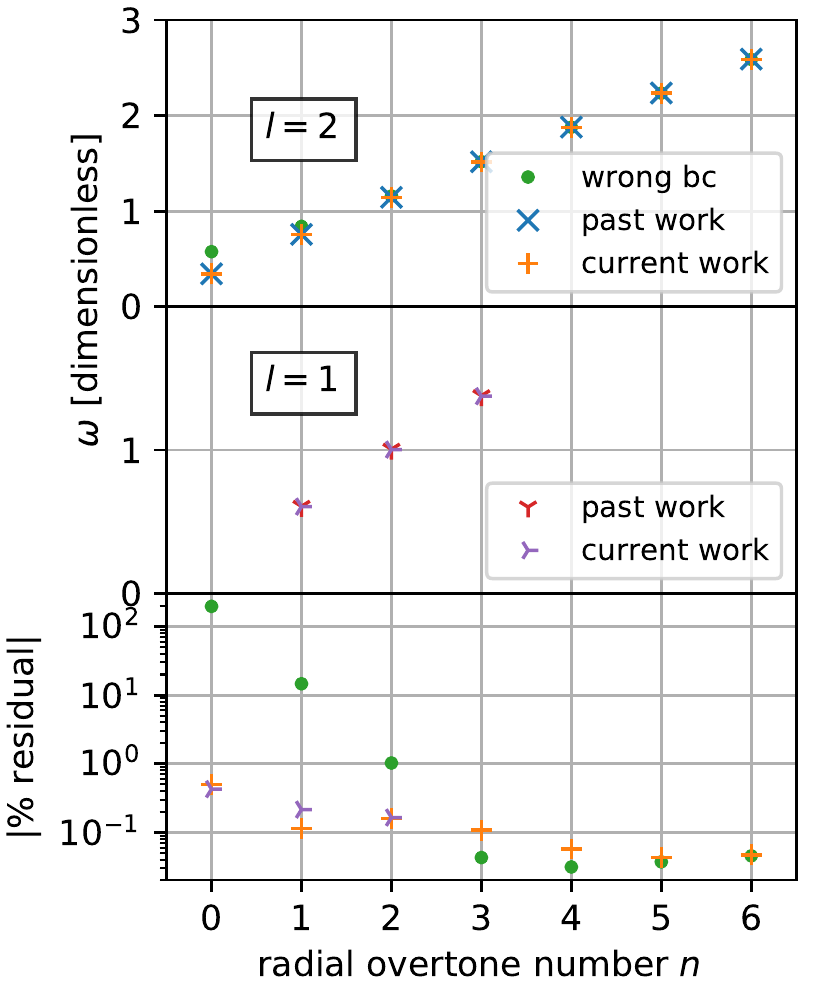}

\caption{Comparison between axisymmetric $l=1$ and $l=2$ mode frequencies obtained in this work versus past work~\cite{horedt2004polytropes} pg.~387, for a $\Gamma = 5/3$ polytrope. The frequencies are displayed in dimensionless form $\omega = \sqrt{\sigma^2/4\pi G \rho_c}$ where $\sigma=2\pi f$ is the angular frequency and $\rho_c$ is the central density. The wrong boundary condition $\delta\hat{\Phi}\vert_{r_0}=0=\partial_r \delta\hat{\Phi}\vert_{r_0}$ (green dots) has a large error for the lower overtones. With the correct boundary conditions (Eqs.~\eqref{eq:origin}), we obtain at worst $\sim 0.4 \%$ residual for the fundamental $n=0$ mode.} \label{fig:newt_test}
\end{figure}

\begin{figure}
\vspace{0.1cm}
\centering
\includegraphics[width=0.48\textwidth]{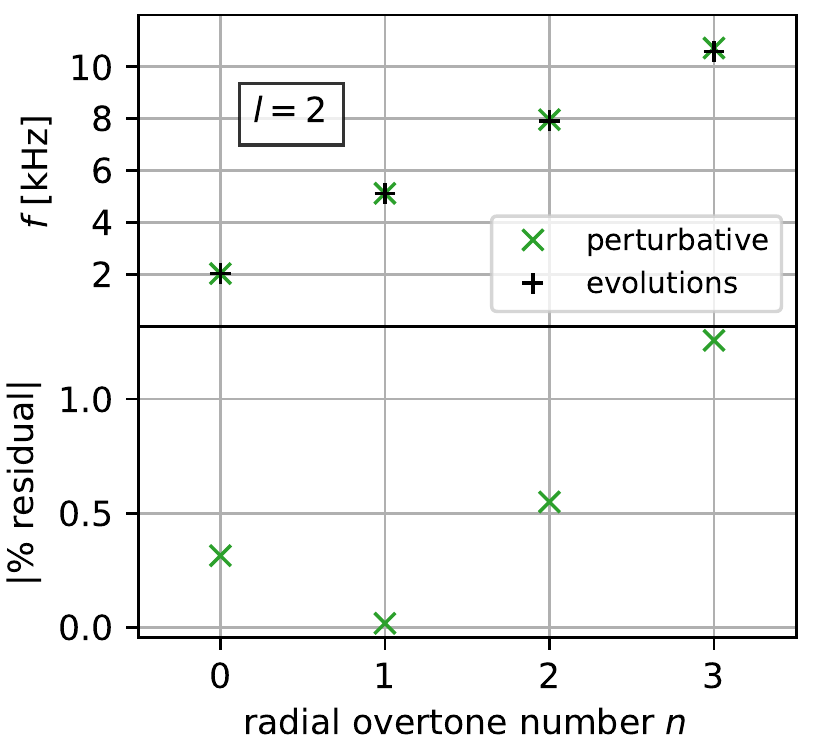}

\caption{Comparison between axisymmetric $l=2$ mode frequencies obtained perturbatively in this work versus using full nonlinear \texttt{FLASH} evolutions in past work~\cite{westernacher2018turbulence,westernacher2019multimessenger}, for a $\Gamma = 2$ polytropic star with $P=\kappa \rho^\Gamma$, $\rho_c = 1.28\times 10^{-3}$, and $\kappa=100$ in geometrized units.} \label{fig:FLASHTOV_test}
\end{figure}

\subsection{Newtonian polytropic star}

Fig.~\ref{fig:newt_test} displays a comparison between $l=1$ and $l=2$ mode frequencies we obtain for a $\Gamma=5/3$ Newtonian polytropic star. The polytropic constant $\kappa$, where $P=\kappa \rho^\Gamma$ is arbitrary, and we display the frequencies in dimensionless form
\begin{eqnarray}
\omega \equiv \sqrt{\frac{\sigma^2}{4 \pi G \rho_c}},
\end{eqnarray}
where $\sigma = 2\pi f$ is the angular frequency and $\rho_c$ is the central rest mass density. We impose a vanishing Lagrangian pressure perturbation at the surface, Eq.~\eqref{eq:DeltaP0}. We terminate the frequency search when the update becomes smaller than $0.5$ Hz (we set the stellar mass to $M=1.4 M_{\odot}$ and radius to $R = 12$ km, yielding mode frequencies $\gtrsim 2$ kHz). The frequencies compare favorably with past work (\cite{horedt2004polytropes} pg.~387 and references therein), except when the outer boundary condition for the Newtonian potential is disregarded (setting $\delta\hat{\Phi} =\partial_r \delta \hat{\Phi}=0$ at the starting point of outward integration), as done in~\cite{morozova2018gravitational} and repeated in subsequent work, including~\cite{westernacher2018turbulence, radice2019characterizing, westernacher2019multimessenger}.

\subsection{\texttt{FLASH} Tolman-Oppenheimer-Volkoff star}

Fig.~\ref{fig:FLASHTOV_test} displays a comparison between $l=2$ modes computed perturbatively in this work with those extracted in~\cite{westernacher2018turbulence, westernacher2019multimessenger} from a fully nonlinear \texttt{FLASH} simulation of an equilibrium $\Gamma=2$ star with $\kappa = 100$ and $\rho_c = 1.28\times 10^{-3}$ in geometrized units. We impose vanishing Lagrangian pressure perturbation at the surface, Eq.~\eqref{eq:DeltaP0}. The frequency search terminates when the update is less than $0.5$ Hz.

This test demonstrates that the non-radial modes of pseudo-Newtonian systems, as simulated in eg.~\texttt{FLASH}~\cite{fryxell2000flash,dubey2009extensible},~\texttt{FORNAX}~\cite{skinner2019fornax},~\texttt{CHIMERA}~\cite{bruenn2018chimera}, are determined by a purely Newtonian perturbative calculation. Radial perturbations of the gravitational potential, which would require knowledge of an equation of motion determining the ``effectively GR'' monopole (\cite{marek2006exploring} Case A), do not arise anywhere when one solves for non-radial modes.

\subsection{CCSN system}

\begin{figure}
\vspace{0.1cm}
\centering
\includegraphics[width=0.48\textwidth]{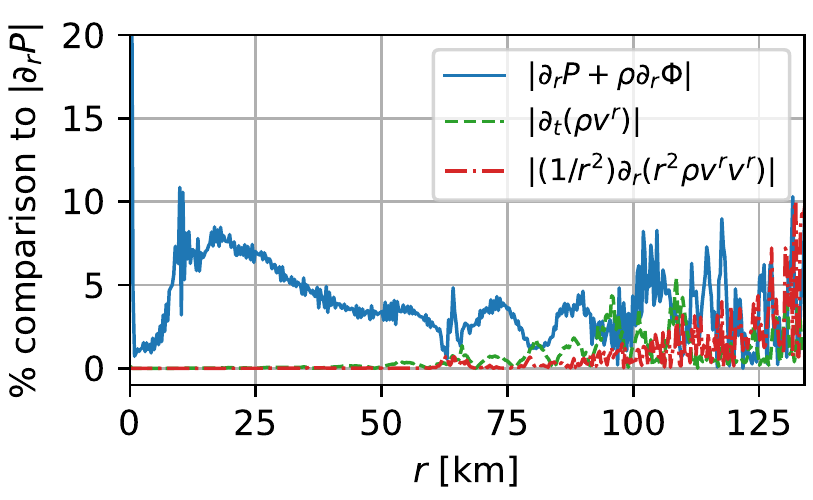}

\caption{A comparison between the magnitude of different terms in the spherically-symmetric Euler equation, as applied to the spherically-averaged snapshot of the CCSN system at $40$ ms. The equilibrium condition $\partial_r P + \rho\partial_r \Phi = 0$ is only satisfied at the $\sim 5\%$ level, which is commensurate with the frequency mismatch between the simulation and the best-fit perturbative solution. The non-equilibrium terms $\partial_t (\rho v^r)$ and $r^{-2} \partial_r (\rho v^r v^r)$ give a negligible contribution to the balance at $r<50$ km (sub-0.1\%), and rises to $\sim 1$\% around $r=100$ km.} \label{fig:bgresidual}
\end{figure}

We know based on the previous tests that the perturbative system is the consistent linearization of the equations of motion being simulated. However, when applying it to the CCSN system, we are dealing with a non-spherical system which we subject to a spherical averaging before performing the perturbative calculation, and it is not in hydrostatic equilibrium. In Fig.~\ref{fig:bgresidual} we compare the magnitude of different terms in the spherically-symmetric Euler equation
\begin{eqnarray}
0= \partial_t \left(\rho v^r\right) + \frac{1}{r^2} \partial_r \left( r^2 \rho v^r v^r \right) + \partial_r P + \rho \partial_r \Phi,
\end{eqnarray}
as a percentage comparison to $\vert \partial_r P\vert$. The equilibrium condition $\partial_r P + \rho \partial_r \Phi$ is satisfied at the $\sim 5$\% level. Note that neutrino pressure gradients should also have a contribution to this balance, but their perturbations would introduce additional equations of motion so we have decided to neglect them. Furthermore, neutrino pressure gradients should gradually decouple from the fluid as one moves away from the PNS center, so introducing them into the background solution requires care. The level of violation of the hydrostatic equilibrium condition should be taken as a cautionary note when applying this perturbative calculation to dynamical systems such as CCSNe.

By comparison, the other terms which encode time-dependence of the background solution ($\partial_t(\rho v^r)$) or its non-steadiness ($v^r=\,$constant$\,\neq 0$) are not large enough to account for the degree of non-equilibrium (sub-0.1\% for $r<50$ km rising to 1\% around $r=100$ km). This suggests that generalizing the perturbative scheme to a time-dependent or unsteady background would not yield significant improvements in the perturbative calculations presented in this work.

\renewcommand\bibname{References}

\bibliographystyle{apsrev4-1}
\bibliography{fluidbib}

\begin{thebibliography}{34}%
\makeatletter
\providecommand \@ifxundefined [1]{%
 \@ifx{#1\undefined}
}%
\providecommand \@ifnum [1]{%
 \ifnum #1\expandafter \@firstoftwo
 \else \expandafter \@secondoftwo
 \fi
}%
\providecommand \@ifx [1]{%
 \ifx #1\expandafter \@firstoftwo
 \else \expandafter \@secondoftwo
 \fi
}%
\providecommand \natexlab [1]{#1}%
\providecommand \enquote  [1]{``#1''}%
\providecommand \bibnamefont  [1]{#1}%
\providecommand \bibfnamefont [1]{#1}%
\providecommand \citenamefont [1]{#1}%
\providecommand \href@noop [0]{\@secondoftwo}%
\providecommand \href [0]{\begingroup \@sanitize@url \@href}%
\providecommand \@href[1]{\@@startlink{#1}\@@href}%
\providecommand \@@href[1]{\endgroup#1\@@endlink}%
\providecommand \@sanitize@url [0]{\catcode `\\12\catcode `\$12\catcode
  `\&12\catcode `\#12\catcode `\^12\catcode `\_12\catcode `\%12\relax}%
\providecommand \@@startlink[1]{}%
\providecommand \@@endlink[0]{}%
\providecommand \url  [0]{\begingroup\@sanitize@url \@url }%
\providecommand \@url [1]{\endgroup\@href {#1}{\urlprefix }}%
\providecommand \urlprefix  [0]{URL }%
\providecommand \Eprint [0]{\href }%
\providecommand \doibase [0]{http://dx.doi.org/}%
\providecommand \selectlanguage [0]{\@gobble}%
\providecommand \bibinfo  [0]{\@secondoftwo}%
\providecommand \bibfield  [0]{\@secondoftwo}%
\providecommand \translation [1]{[#1]}%
\providecommand \BibitemOpen [0]{}%
\providecommand \bibitemStop [0]{}%
\providecommand \bibitemNoStop [0]{.\EOS\space}%
\providecommand \EOS [0]{\spacefactor3000\relax}%
\providecommand \BibitemShut  [1]{\csname bibitem#1\endcsname}%
\let\auto@bib@innerbib\@empty
\bibitem [{\citenamefont {Murphy}\ \emph {et~al.}(2009)\citenamefont {Murphy},
  \citenamefont {Ott},\ and\ \citenamefont {Burrows}}]{murphy2009model}%
  \BibitemOpen
  \bibfield  {author} {\bibinfo {author} {\bibfnamefont {J.~W.}\ \bibnamefont
  {Murphy}}, \bibinfo {author} {\bibfnamefont {C.~D.}\ \bibnamefont {Ott}}, \
  and\ \bibinfo {author} {\bibfnamefont {A.}~\bibnamefont {Burrows}},\
  }\href@noop {} {\bibfield  {journal} {\bibinfo  {journal} {The Astrophysical
  Journal}\ }\textbf {\bibinfo {volume} {707}},\ \bibinfo {pages} {1173}
  (\bibinfo {year} {2009})}\BibitemShut {NoStop}%
\bibitem [{\citenamefont {M{\"u}ller}\ \emph {et~al.}(2013)\citenamefont
  {M{\"u}ller}, \citenamefont {Janka},\ and\ \citenamefont
  {Marek}}]{muller2013new}%
  \BibitemOpen
  \bibfield  {author} {\bibinfo {author} {\bibfnamefont {B.}~\bibnamefont
  {M{\"u}ller}}, \bibinfo {author} {\bibfnamefont {H.-T.}\ \bibnamefont
  {Janka}}, \ and\ \bibinfo {author} {\bibfnamefont {A.}~\bibnamefont
  {Marek}},\ }\href@noop {} {\bibfield  {journal} {\bibinfo  {journal} {The
  Astrophysical Journal}\ }\textbf {\bibinfo {volume} {766}},\ \bibinfo {pages}
  {43} (\bibinfo {year} {2013})}\BibitemShut {NoStop}%
\bibitem [{\citenamefont {Cerd{\'a}-Dur{\'a}n}\ \emph
  {et~al.}(2013)\citenamefont {Cerd{\'a}-Dur{\'a}n}, \citenamefont {DeBrye},
  \citenamefont {Aloy}, \citenamefont {Font},\ and\ \citenamefont
  {Obergaulinger}}]{cerda2013gravitational}%
  \BibitemOpen
  \bibfield  {author} {\bibinfo {author} {\bibfnamefont {P.}~\bibnamefont
  {Cerd{\'a}-Dur{\'a}n}}, \bibinfo {author} {\bibfnamefont {N.}~\bibnamefont
  {DeBrye}}, \bibinfo {author} {\bibfnamefont {M.~A.}\ \bibnamefont {Aloy}},
  \bibinfo {author} {\bibfnamefont {J.~A.}\ \bibnamefont {Font}}, \ and\
  \bibinfo {author} {\bibfnamefont {M.}~\bibnamefont {Obergaulinger}},\
  }\href@noop {} {\bibfield  {journal} {\bibinfo  {journal} {The Astrophysical
  Journal Letters}\ }\textbf {\bibinfo {volume} {779}},\ \bibinfo {pages} {L18}
  (\bibinfo {year} {2013})}\BibitemShut {NoStop}%
\bibitem [{\citenamefont {Fuller}\ \emph {et~al.}(2015)\citenamefont {Fuller},
  \citenamefont {Klion}, \citenamefont {Abdikamalov},\ and\ \citenamefont
  {Ott}}]{fuller2015supernova}%
  \BibitemOpen
  \bibfield  {author} {\bibinfo {author} {\bibfnamefont {J.}~\bibnamefont
  {Fuller}}, \bibinfo {author} {\bibfnamefont {H.}~\bibnamefont {Klion}},
  \bibinfo {author} {\bibfnamefont {E.}~\bibnamefont {Abdikamalov}}, \ and\
  \bibinfo {author} {\bibfnamefont {C.~D.}\ \bibnamefont {Ott}},\ }\href@noop
  {} {\bibfield  {journal} {\bibinfo  {journal} {Monthly Notices of the Royal
  Astronomical Society}\ }\textbf {\bibinfo {volume} {450}},\ \bibinfo {pages}
  {414} (\bibinfo {year} {2015})}\BibitemShut {NoStop}%
\bibitem [{\citenamefont {Torres-Forn{\'e}}\ \emph {et~al.}(2017)\citenamefont
  {Torres-Forn{\'e}}, \citenamefont {Cerd{\'a}-Dur{\'a}n}, \citenamefont
  {Passamonti},\ and\ \citenamefont {Font}}]{torres2017towards}%
  \BibitemOpen
  \bibfield  {author} {\bibinfo {author} {\bibfnamefont {A.}~\bibnamefont
  {Torres-Forn{\'e}}}, \bibinfo {author} {\bibfnamefont {P.}~\bibnamefont
  {Cerd{\'a}-Dur{\'a}n}}, \bibinfo {author} {\bibfnamefont {A.}~\bibnamefont
  {Passamonti}}, \ and\ \bibinfo {author} {\bibfnamefont {J.~A.}\ \bibnamefont
  {Font}},\ }\href@noop {} {\bibfield  {journal} {\bibinfo  {journal} {Monthly
  Notices of the Royal Astronomical Society}\ }\textbf {\bibinfo {volume}
  {474}},\ \bibinfo {pages} {5272} (\bibinfo {year} {2017})}\BibitemShut
  {NoStop}%
\bibitem [{\citenamefont {Morozova}\ \emph {et~al.}(2018)\citenamefont
  {Morozova}, \citenamefont {Radice}, \citenamefont {Burrows},\ and\
  \citenamefont {Vartanyan}}]{morozova2018gravitational}%
  \BibitemOpen
  \bibfield  {author} {\bibinfo {author} {\bibfnamefont {V.}~\bibnamefont
  {Morozova}}, \bibinfo {author} {\bibfnamefont {D.}~\bibnamefont {Radice}},
  \bibinfo {author} {\bibfnamefont {A.}~\bibnamefont {Burrows}}, \ and\
  \bibinfo {author} {\bibfnamefont {D.}~\bibnamefont {Vartanyan}},\ }\href@noop
  {} {\bibfield  {journal} {\bibinfo  {journal} {The Astrophysical Journal}\
  }\textbf {\bibinfo {volume} {861}},\ \bibinfo {pages} {10} (\bibinfo {year}
  {2018})}\BibitemShut {NoStop}%
\bibitem [{\citenamefont {Torres-Forn{\'e}}\ \emph {et~al.}(2018)\citenamefont
  {Torres-Forn{\'e}}, \citenamefont {Cerd{\'a}-Dur{\'a}n}, \citenamefont
  {Passamonti}, \citenamefont {Obergaulinger},\ and\ \citenamefont
  {Font}}]{torres2018towards}%
  \BibitemOpen
  \bibfield  {author} {\bibinfo {author} {\bibfnamefont {A.}~\bibnamefont
  {Torres-Forn{\'e}}}, \bibinfo {author} {\bibfnamefont {P.}~\bibnamefont
  {Cerd{\'a}-Dur{\'a}n}}, \bibinfo {author} {\bibfnamefont {A.}~\bibnamefont
  {Passamonti}}, \bibinfo {author} {\bibfnamefont {M.}~\bibnamefont
  {Obergaulinger}}, \ and\ \bibinfo {author} {\bibfnamefont {J.~A.}\
  \bibnamefont {Font}},\ }\href@noop {} {\bibfield  {journal} {\bibinfo
  {journal} {Monthly Notices of the Royal Astronomical Society}\ }\textbf
  {\bibinfo {volume} {482}},\ \bibinfo {pages} {3967} (\bibinfo {year}
  {2018})}\BibitemShut {NoStop}%
\bibitem [{\citenamefont
  {Westernacher-Schneider}(2018)}]{westernacher2018turbulence}%
  \BibitemOpen
  \bibfield  {author} {\bibinfo {author} {\bibfnamefont {J.~R.}\ \bibnamefont
  {Westernacher-Schneider}},\ }\emph {\bibinfo {title} {Turbulence, Gravity,
  and Multimessenger Asteroseismology}},\ \href@noop {} {Ph.D. thesis}
  (\bibinfo {year} {2018})\BibitemShut {NoStop}%
\bibitem [{\citenamefont {Torres-Forn{\'e}}\ \emph {et~al.}(2019)\citenamefont
  {Torres-Forn{\'e}}, \citenamefont {Cerd{\'a}-Dur{\'a}n}, \citenamefont
  {Obergaulinger}, \citenamefont {Muller},\ and\ \citenamefont
  {Font}}]{torres2019universal}%
  \BibitemOpen
  \bibfield  {author} {\bibinfo {author} {\bibfnamefont {A.}~\bibnamefont
  {Torres-Forn{\'e}}}, \bibinfo {author} {\bibfnamefont {P.}~\bibnamefont
  {Cerd{\'a}-Dur{\'a}n}}, \bibinfo {author} {\bibfnamefont {M.}~\bibnamefont
  {Obergaulinger}}, \bibinfo {author} {\bibfnamefont {B.}~\bibnamefont
  {Muller}}, \ and\ \bibinfo {author} {\bibfnamefont {J.~A.}\ \bibnamefont
  {Font}},\ }\href@noop {} {\bibfield  {journal} {\bibinfo  {journal} {arXiv
  preprint arXiv:1902.10048}\ } (\bibinfo {year} {2019})}\BibitemShut {NoStop}%
\bibitem [{\citenamefont {Vartanyan}\ \emph {et~al.}(2019)\citenamefont
  {Vartanyan}, \citenamefont {Burrows},\ and\ \citenamefont
  {Radice}}]{vartanyan2019temporal}%
  \BibitemOpen
  \bibfield  {author} {\bibinfo {author} {\bibfnamefont {D.}~\bibnamefont
  {Vartanyan}}, \bibinfo {author} {\bibfnamefont {A.}~\bibnamefont {Burrows}},
  \ and\ \bibinfo {author} {\bibfnamefont {D.}~\bibnamefont {Radice}},\
  }\href@noop {} {\bibfield  {journal} {\bibinfo  {journal} {arXiv preprint
  arXiv:1906.08787}\ } (\bibinfo {year} {2019})}\BibitemShut {NoStop}%
\bibitem [{\citenamefont {Sotani}\ \emph {et~al.}(2019)\citenamefont {Sotani},
  \citenamefont {Kuroda}, \citenamefont {Takiwaki},\ and\ \citenamefont
  {Kotake}}]{sotani2019dependence}%
  \BibitemOpen
  \bibfield  {author} {\bibinfo {author} {\bibfnamefont {H.}~\bibnamefont
  {Sotani}}, \bibinfo {author} {\bibfnamefont {T.}~\bibnamefont {Kuroda}},
  \bibinfo {author} {\bibfnamefont {T.}~\bibnamefont {Takiwaki}}, \ and\
  \bibinfo {author} {\bibfnamefont {K.}~\bibnamefont {Kotake}},\ }\href@noop {}
  {\bibfield  {journal} {\bibinfo  {journal} {arXiv preprint arXiv:1906.04354}\
  } (\bibinfo {year} {2019})}\BibitemShut {NoStop}%
\bibitem [{\citenamefont {Westernacher-Schneider}\ \emph
  {et~al.}(2019)\citenamefont {Westernacher-Schneider}, \citenamefont
  {O'Connor}, \citenamefont {O'Sullivan}, \citenamefont {Tamborra},
  \citenamefont {Wu}, \citenamefont {Couch},\ and\ \citenamefont
  {Malmenbeck}}]{westernacher2019multimessenger}%
  \BibitemOpen
  \bibfield  {author} {\bibinfo {author} {\bibfnamefont {J.~R.}\ \bibnamefont
  {Westernacher-Schneider}}, \bibinfo {author} {\bibfnamefont {E.}~\bibnamefont
  {O'Connor}}, \bibinfo {author} {\bibfnamefont {E.}~\bibnamefont
  {O'Sullivan}}, \bibinfo {author} {\bibfnamefont {I.}~\bibnamefont
  {Tamborra}}, \bibinfo {author} {\bibfnamefont {M.-R.}\ \bibnamefont {Wu}},
  \bibinfo {author} {\bibfnamefont {S.~M.}\ \bibnamefont {Couch}}, \ and\
  \bibinfo {author} {\bibfnamefont {F.}~\bibnamefont {Malmenbeck}},\
  }\href@noop {} {\bibfield  {journal} {\bibinfo  {journal} {Physical Review
  D}\ }\textbf {\bibinfo {volume} {100}},\ \bibinfo {pages} {123009} (\bibinfo
  {year} {2019})}\BibitemShut {NoStop}%
\bibitem [{\citenamefont {Warren}\ \emph {et~al.}(2019)\citenamefont {Warren},
  \citenamefont {Couch}, \citenamefont {O'Connor},\ and\ \citenamefont
  {Morozova}}]{warren2019constraining}%
  \BibitemOpen
  \bibfield  {author} {\bibinfo {author} {\bibfnamefont {M.~L.}\ \bibnamefont
  {Warren}}, \bibinfo {author} {\bibfnamefont {S.~M.}\ \bibnamefont {Couch}},
  \bibinfo {author} {\bibfnamefont {E.~P.}\ \bibnamefont {O'Connor}}, \ and\
  \bibinfo {author} {\bibfnamefont {V.}~\bibnamefont {Morozova}},\ }\href@noop
  {} {\bibfield  {journal} {\bibinfo  {journal} {arXiv preprint
  arXiv:1912.03328}\ } (\bibinfo {year} {2019})}\BibitemShut {NoStop}%
\bibitem [{\citenamefont {Rampp}\ and\ \citenamefont
  {Janka}(2002)}]{rampp2002radiation}%
  \BibitemOpen
  \bibfield  {author} {\bibinfo {author} {\bibfnamefont {M.}~\bibnamefont
  {Rampp}}\ and\ \bibinfo {author} {\bibfnamefont {H.-T.}\ \bibnamefont
  {Janka}},\ }\href@noop {} {\bibfield  {journal} {\bibinfo  {journal}
  {Astronomy \& Astrophysics}\ }\textbf {\bibinfo {volume} {396}},\ \bibinfo
  {pages} {361} (\bibinfo {year} {2002})}\BibitemShut {NoStop}%
\bibitem [{\citenamefont {M{\"u}ller}\ \emph {et~al.}(2010)\citenamefont
  {M{\"u}ller}, \citenamefont {Janka},\ and\ \citenamefont
  {Dimmelmeier}}]{muller2010new}%
  \BibitemOpen
  \bibfield  {author} {\bibinfo {author} {\bibfnamefont {B.}~\bibnamefont
  {M{\"u}ller}}, \bibinfo {author} {\bibfnamefont {H.-T.}\ \bibnamefont
  {Janka}}, \ and\ \bibinfo {author} {\bibfnamefont {H.}~\bibnamefont
  {Dimmelmeier}},\ }\href@noop {} {\bibfield  {journal} {\bibinfo  {journal}
  {The Astrophysical Journal Supplement Series}\ }\textbf {\bibinfo {volume}
  {189}},\ \bibinfo {pages} {104} (\bibinfo {year} {2010})}\BibitemShut
  {NoStop}%
\bibitem [{\citenamefont {M{\"u}ller}\ \emph {et~al.}(2012)\citenamefont
  {M{\"u}ller}, \citenamefont {Janka},\ and\ \citenamefont
  {Marek}}]{muller2012new}%
  \BibitemOpen
  \bibfield  {author} {\bibinfo {author} {\bibfnamefont {B.}~\bibnamefont
  {M{\"u}ller}}, \bibinfo {author} {\bibfnamefont {H.-T.}\ \bibnamefont
  {Janka}}, \ and\ \bibinfo {author} {\bibfnamefont {A.}~\bibnamefont
  {Marek}},\ }\href@noop {} {\bibfield  {journal} {\bibinfo  {journal} {The
  Astrophysical Journal}\ }\textbf {\bibinfo {volume} {756}},\ \bibinfo {pages}
  {84} (\bibinfo {year} {2012})}\BibitemShut {NoStop}%
\bibitem [{\citenamefont {M{\"u}ller}\ and\ \citenamefont
  {Janka}(2014)}]{muller2014new}%
  \BibitemOpen
  \bibfield  {author} {\bibinfo {author} {\bibfnamefont {B.}~\bibnamefont
  {M{\"u}ller}}\ and\ \bibinfo {author} {\bibfnamefont {H.-T.}\ \bibnamefont
  {Janka}},\ }\href@noop {} {\bibfield  {journal} {\bibinfo  {journal} {The
  Astrophysical Journal}\ }\textbf {\bibinfo {volume} {788}},\ \bibinfo {pages}
  {82} (\bibinfo {year} {2014})}\BibitemShut {NoStop}%
\bibitem [{\citenamefont {Fryxell}\ \emph {et~al.}(2000)\citenamefont
  {Fryxell}, \citenamefont {Olson}, \citenamefont {Ricker}, \citenamefont
  {Timmes}, \citenamefont {Zingale}, \citenamefont {Lamb}, \citenamefont
  {MacNeice}, \citenamefont {Rosner}, \citenamefont {Truran},\ and\
  \citenamefont {Tufo}}]{fryxell2000flash}%
  \BibitemOpen
  \bibfield  {author} {\bibinfo {author} {\bibfnamefont {B.}~\bibnamefont
  {Fryxell}}, \bibinfo {author} {\bibfnamefont {K.}~\bibnamefont {Olson}},
  \bibinfo {author} {\bibfnamefont {P.}~\bibnamefont {Ricker}}, \bibinfo
  {author} {\bibfnamefont {F.}~\bibnamefont {Timmes}}, \bibinfo {author}
  {\bibfnamefont {M.}~\bibnamefont {Zingale}}, \bibinfo {author} {\bibfnamefont
  {D.}~\bibnamefont {Lamb}}, \bibinfo {author} {\bibfnamefont {P.}~\bibnamefont
  {MacNeice}}, \bibinfo {author} {\bibfnamefont {R.}~\bibnamefont {Rosner}},
  \bibinfo {author} {\bibfnamefont {J.}~\bibnamefont {Truran}}, \ and\ \bibinfo
  {author} {\bibfnamefont {H.}~\bibnamefont {Tufo}},\ }\href@noop {} {\bibfield
   {journal} {\bibinfo  {journal} {The Astrophysical Journal Supplement
  Series}\ }\textbf {\bibinfo {volume} {131}},\ \bibinfo {pages} {273}
  (\bibinfo {year} {2000})}\BibitemShut {NoStop}%
\bibitem [{\citenamefont {Dubey}\ \emph {et~al.}(2009)\citenamefont {Dubey},
  \citenamefont {Antypas}, \citenamefont {Ganapathy}, \citenamefont {Reid},
  \citenamefont {Riley}, \citenamefont {Sheeler}, \citenamefont {Siegel},\ and\
  \citenamefont {Weide}}]{dubey2009extensible}%
  \BibitemOpen
  \bibfield  {author} {\bibinfo {author} {\bibfnamefont {A.}~\bibnamefont
  {Dubey}}, \bibinfo {author} {\bibfnamefont {K.}~\bibnamefont {Antypas}},
  \bibinfo {author} {\bibfnamefont {M.~K.}\ \bibnamefont {Ganapathy}}, \bibinfo
  {author} {\bibfnamefont {L.~B.}\ \bibnamefont {Reid}}, \bibinfo {author}
  {\bibfnamefont {K.}~\bibnamefont {Riley}}, \bibinfo {author} {\bibfnamefont
  {D.}~\bibnamefont {Sheeler}}, \bibinfo {author} {\bibfnamefont
  {A.}~\bibnamefont {Siegel}}, \ and\ \bibinfo {author} {\bibfnamefont
  {K.}~\bibnamefont {Weide}},\ }\href@noop {} {\bibfield  {journal} {\bibinfo
  {journal} {Parallel Computing}\ }\textbf {\bibinfo {volume} {35}},\ \bibinfo
  {pages} {512} (\bibinfo {year} {2009})}\BibitemShut {NoStop}%
\bibitem [{\citenamefont {Skinner}\ \emph {et~al.}(2019)\citenamefont
  {Skinner}, \citenamefont {Dolence}, \citenamefont {Burrows}, \citenamefont
  {Radice},\ and\ \citenamefont {Vartanyan}}]{skinner2019fornax}%
  \BibitemOpen
  \bibfield  {author} {\bibinfo {author} {\bibfnamefont {M.~A.}\ \bibnamefont
  {Skinner}}, \bibinfo {author} {\bibfnamefont {J.~C.}\ \bibnamefont
  {Dolence}}, \bibinfo {author} {\bibfnamefont {A.}~\bibnamefont {Burrows}},
  \bibinfo {author} {\bibfnamefont {D.}~\bibnamefont {Radice}}, \ and\ \bibinfo
  {author} {\bibfnamefont {D.}~\bibnamefont {Vartanyan}},\ }\href@noop {}
  {\bibfield  {journal} {\bibinfo  {journal} {The Astrophysical Journal
  Supplement Series}\ }\textbf {\bibinfo {volume} {241}},\ \bibinfo {pages} {7}
  (\bibinfo {year} {2019})}\BibitemShut {NoStop}%
\bibitem [{\citenamefont {Bruenn}\ \emph {et~al.}(2018)\citenamefont {Bruenn},
  \citenamefont {Blondin}, \citenamefont {Hix}, \citenamefont {Lentz},
  \citenamefont {Messer}, \citenamefont {Mezzacappa}, \citenamefont {Endeve},
  \citenamefont {Harris}, \citenamefont {Marronetti}, \citenamefont {Budiardja}
  \emph {et~al.}}]{bruenn2018chimera}%
  \BibitemOpen
  \bibfield  {author} {\bibinfo {author} {\bibfnamefont {S.~W.}\ \bibnamefont
  {Bruenn}}, \bibinfo {author} {\bibfnamefont {J.~M.}\ \bibnamefont {Blondin}},
  \bibinfo {author} {\bibfnamefont {W.~R.}\ \bibnamefont {Hix}}, \bibinfo
  {author} {\bibfnamefont {E.~J.}\ \bibnamefont {Lentz}}, \bibinfo {author}
  {\bibfnamefont {O.}~\bibnamefont {Messer}}, \bibinfo {author} {\bibfnamefont
  {A.}~\bibnamefont {Mezzacappa}}, \bibinfo {author} {\bibfnamefont
  {E.}~\bibnamefont {Endeve}}, \bibinfo {author} {\bibfnamefont {J.~A.}\
  \bibnamefont {Harris}}, \bibinfo {author} {\bibfnamefont {P.}~\bibnamefont
  {Marronetti}}, \bibinfo {author} {\bibfnamefont {R.~D.}\ \bibnamefont
  {Budiardja}},  \emph {et~al.},\ }\href@noop {} {\bibfield  {journal}
  {\bibinfo  {journal} {arXiv preprint arXiv:1809.05608}\ } (\bibinfo {year}
  {2018})}\BibitemShut {NoStop}%
\bibitem [{\citenamefont {Mueller}\ \emph {et~al.}(2008)\citenamefont
  {Mueller}, \citenamefont {Dimmelmeier},\ and\ \citenamefont
  {Mueller}}]{mueller2008exploring}%
  \BibitemOpen
  \bibfield  {author} {\bibinfo {author} {\bibfnamefont {B.}~\bibnamefont
  {Mueller}}, \bibinfo {author} {\bibfnamefont {H.}~\bibnamefont
  {Dimmelmeier}}, \ and\ \bibinfo {author} {\bibfnamefont {E.}~\bibnamefont
  {Mueller}},\ }\href@noop {} {\bibfield  {journal} {\bibinfo  {journal}
  {Astronomy \& Astrophysics}\ }\textbf {\bibinfo {volume} {489}},\ \bibinfo
  {pages} {301} (\bibinfo {year} {2008})}\BibitemShut {NoStop}%
\bibitem [{\citenamefont {Poisson}\ and\ \citenamefont
  {Will}(2014)}]{poisson2014gravity}%
  \BibitemOpen
  \bibfield  {author} {\bibinfo {author} {\bibfnamefont {E.}~\bibnamefont
  {Poisson}}\ and\ \bibinfo {author} {\bibfnamefont {C.~M.}\ \bibnamefont
  {Will}},\ }\href@noop {} {\emph {\bibinfo {title} {Gravity: Newtonian,
  post-newtonian, relativistic}}}\ (\bibinfo  {publisher} {Cambridge University
  Press},\ \bibinfo {year} {2014})\BibitemShut {NoStop}%
\bibitem [{\citenamefont {Christensen-Dalsgaard}\ \emph
  {et~al.}(1991)\citenamefont {Christensen-Dalsgaard}, \citenamefont
  {Berthomieu}, \citenamefont {Cox}, \citenamefont {Livingston},\ and\
  \citenamefont {Matthews}}]{christensen1991solar}%
  \BibitemOpen
  \bibfield  {author} {\bibinfo {author} {\bibfnamefont {J.}~\bibnamefont
  {Christensen-Dalsgaard}}, \bibinfo {author} {\bibfnamefont {G.}~\bibnamefont
  {Berthomieu}}, \bibinfo {author} {\bibfnamefont {A.}~\bibnamefont {Cox}},
  \bibinfo {author} {\bibfnamefont {W.}~\bibnamefont {Livingston}}, \ and\
  \bibinfo {author} {\bibfnamefont {M.}~\bibnamefont {Matthews}},\ }\href@noop
  {} {\bibfield  {journal} {\bibinfo  {journal} {The University of Arizona
  Press, Tucson}\ }\textbf {\bibinfo {volume} {401}} (\bibinfo {year}
  {1991})}\BibitemShut {NoStop}%
\bibitem [{\citenamefont {Radice}\ \emph {et~al.}(2019)\citenamefont {Radice},
  \citenamefont {Morozova}, \citenamefont {Burrows}, \citenamefont
  {Vartanyan},\ and\ \citenamefont {Nagakura}}]{radice2019characterizing}%
  \BibitemOpen
  \bibfield  {author} {\bibinfo {author} {\bibfnamefont {D.}~\bibnamefont
  {Radice}}, \bibinfo {author} {\bibfnamefont {V.}~\bibnamefont {Morozova}},
  \bibinfo {author} {\bibfnamefont {A.}~\bibnamefont {Burrows}}, \bibinfo
  {author} {\bibfnamefont {D.}~\bibnamefont {Vartanyan}}, \ and\ \bibinfo
  {author} {\bibfnamefont {H.}~\bibnamefont {Nagakura}},\ }\href@noop {}
  {\bibfield  {journal} {\bibinfo  {journal} {The Astrophysical Journal
  Letters}\ }\textbf {\bibinfo {volume} {876}},\ \bibinfo {pages} {L9}
  (\bibinfo {year} {2019})}\BibitemShut {NoStop}%
\bibitem [{\citenamefont {Marek}\ \emph {et~al.}(2006)\citenamefont {Marek},
  \citenamefont {Dimmelmeier}, \citenamefont {Janka}, \citenamefont
  {M{\"u}ller},\ and\ \citenamefont {Buras}}]{marek2006exploring}%
  \BibitemOpen
  \bibfield  {author} {\bibinfo {author} {\bibfnamefont {A.}~\bibnamefont
  {Marek}}, \bibinfo {author} {\bibfnamefont {H.}~\bibnamefont {Dimmelmeier}},
  \bibinfo {author} {\bibfnamefont {H.-T.}\ \bibnamefont {Janka}}, \bibinfo
  {author} {\bibfnamefont {E.}~\bibnamefont {M{\"u}ller}}, \ and\ \bibinfo
  {author} {\bibfnamefont {R.}~\bibnamefont {Buras}},\ }\href@noop {}
  {\bibfield  {journal} {\bibinfo  {journal} {Astronomy \& Astrophysics}\
  }\textbf {\bibinfo {volume} {445}},\ \bibinfo {pages} {273} (\bibinfo {year}
  {2006})}\BibitemShut {NoStop}%
\bibitem [{\citenamefont {Foglizzo}\ \emph {et~al.}(2007)\citenamefont
  {Foglizzo}, \citenamefont {Galletti}, \citenamefont {Scheck},\ and\
  \citenamefont {Janka}}]{foglizzo2007instability}%
  \BibitemOpen
  \bibfield  {author} {\bibinfo {author} {\bibfnamefont {T.}~\bibnamefont
  {Foglizzo}}, \bibinfo {author} {\bibfnamefont {P.}~\bibnamefont {Galletti}},
  \bibinfo {author} {\bibfnamefont {L.}~\bibnamefont {Scheck}}, \ and\ \bibinfo
  {author} {\bibfnamefont {H.-T.}\ \bibnamefont {Janka}},\ }\href@noop {}
  {\bibfield  {journal} {\bibinfo  {journal} {The Astrophysical Journal}\
  }\textbf {\bibinfo {volume} {654}},\ \bibinfo {pages} {1006} (\bibinfo {year}
  {2007})}\BibitemShut {NoStop}%
\bibitem [{\citenamefont {Laming}(2007)}]{laming2007analytic}%
  \BibitemOpen
  \bibfield  {author} {\bibinfo {author} {\bibfnamefont {J.~M.}\ \bibnamefont
  {Laming}},\ }\href@noop {} {\bibfield  {journal} {\bibinfo  {journal} {The
  Astrophysical Journal}\ }\textbf {\bibinfo {volume} {659}},\ \bibinfo {pages}
  {1449} (\bibinfo {year} {2007})}\BibitemShut {NoStop}%
\bibitem [{\citenamefont {Hunter}(2007)}]{Hunter:2007}%
  \BibitemOpen
  \bibfield  {author} {\bibinfo {author} {\bibfnamefont {J.~D.}\ \bibnamefont
  {Hunter}},\ }\href {\doibase 10.1109/MCSE.2007.55} {\bibfield  {journal}
  {\bibinfo  {journal} {Computing in Science \& Engineering}\ }\textbf
  {\bibinfo {volume} {9}},\ \bibinfo {pages} {90} (\bibinfo {year}
  {2007})}\BibitemShut {NoStop}%
\bibitem [{\citenamefont {{Couch}}(2013)}]{couch13}%
  \BibitemOpen
  \bibfield  {author} {\bibinfo {author} {\bibfnamefont {S.~M.}\ \bibnamefont
  {{Couch}}},\ }\href {\doibase 10.1088/0004-637X/765/1/29} {\bibfield
  {journal} {\bibinfo  {journal} {\apj}\ }\textbf {\bibinfo {volume} {765}},\
  \bibinfo {eid} {29} (\bibinfo {year} {2013})},\ \Eprint
  {http://arxiv.org/abs/1206.4724} {arXiv:1206.4724 [astro-ph.HE]} \BibitemShut
  {NoStop}%
\bibitem [{\citenamefont {{O'Connor}}\ and\ \citenamefont
  {Couch}(2018)}]{o2018two}%
  \BibitemOpen
  \bibfield  {author} {\bibinfo {author} {\bibfnamefont {E.~P.}\ \bibnamefont
  {{O'Connor}}}\ and\ \bibinfo {author} {\bibfnamefont {S.~M.}\ \bibnamefont
  {Couch}},\ }\href@noop {} {\bibfield  {journal} {\bibinfo  {journal} {The
  Astrophysical Journal}\ }\textbf {\bibinfo {volume} {854}},\ \bibinfo {pages}
  {63} (\bibinfo {year} {2018})}\BibitemShut {NoStop}%
\bibitem [{\citenamefont {Jones}\ \emph {et~al.}(01  )\citenamefont {Jones},
  \citenamefont {Oliphant}, \citenamefont {Peterson} \emph {et~al.}}]{scipy}%
  \BibitemOpen
  \bibfield  {author} {\bibinfo {author} {\bibfnamefont {E.}~\bibnamefont
  {Jones}}, \bibinfo {author} {\bibfnamefont {T.}~\bibnamefont {Oliphant}},
  \bibinfo {author} {\bibfnamefont {P.}~\bibnamefont {Peterson}},  \emph
  {et~al.},\ }\href {http://www.scipy.org/} {\enquote {\bibinfo {title}
  {{SciPy}: Open source scientific tools for {Python}},}\ } (\bibinfo {year}
  {2001--}),\ \bibinfo {note} {[Online; accessed 2019-06-01]}\BibitemShut
  {NoStop}%
\bibitem [{\citenamefont {Hurley}\ \emph {et~al.}(1966)\citenamefont {Hurley},
  \citenamefont {Roberts},\ and\ \citenamefont
  {Wright}}]{hurley1966oscillations}%
  \BibitemOpen
  \bibfield  {author} {\bibinfo {author} {\bibfnamefont {M.}~\bibnamefont
  {Hurley}}, \bibinfo {author} {\bibfnamefont {P.}~\bibnamefont {Roberts}}, \
  and\ \bibinfo {author} {\bibfnamefont {K.}~\bibnamefont {Wright}},\
  }\href@noop {} {\bibfield  {journal} {\bibinfo  {journal} {The Astrophysical
  Journal}\ }\textbf {\bibinfo {volume} {143}},\ \bibinfo {pages} {535}
  (\bibinfo {year} {1966})}\BibitemShut {NoStop}%
\bibitem [{\citenamefont {Horedt}(2004)}]{horedt2004polytropes}%
  \BibitemOpen
  \bibfield  {author} {\bibinfo {author} {\bibfnamefont {G.~P.}\ \bibnamefont
  {Horedt}},\ }\href@noop {} {\emph {\bibinfo {title} {Polytropes: applications
  in astrophysics and related fields}}},\ Vol.\ \bibinfo {volume} {306}\
  (\bibinfo  {publisher} {Springer Science \& Business Media},\ \bibinfo {year}
  {2004})\BibitemShut {NoStop}%
\end{thebibliography}%

\end{document}